# The FENIKS Survey: Spectroscopic Confirmation of Massive Quiescent Galaxies at z~3−5

Jacqueline Antwi-Danso [1,2,3,4,5,*] Casey Papovich [1,2] James Esdaile [6] Themiya Nanayakkara [6] Karl Glazebrook [7] Taylor A. Hutchison [8,†] Katherine E. Whitaker [3,9] Z. Cemile Marsan [10] Ruben J. Diaz [11,12] Danilo Marchesini [13] Adam Muzzin [10] Kim-Vy H. Tran [14,15] David J. Setton [16] Yasha Kaushal [17] Joshua S. Speagle (沈佳士) [4,18,5,19] AND Justin Cole [1,2]

[1] *George P. and Cynthia Woods Mitchell Institute for Fundamental Physics and Astronomy, Texas A&M University, College Station, TX 78743, USA*
[2] *Department of Physics and Astronomy, Texas A&M University, 4242 TAMU, College Station, TX 78743, USA*
[3] *Department of Astronomy, University of Massachusetts, Amherst, MA 01003, USA*
[4] *David A. Dunlap Department of Astronomy & Astrophysics, University of Toronto, 50 St George Street, Toronto, ON M5S 3H4, Canada*
[5] *Dunlap Institute for Astronomy & Astrophysics, University of Toronto, 50 St George Street, Toronto, ON M5S 3H4, Canada*
[6] *Centre for Astrophysics and Supercomputing, Swinburne University of Technology, PO Box 218, Hawthorn, VIC 3122, Australia*
[7] *Center for Astrophysics & Supercomputing, Swinburne University, Hawthorn, Victoria, Australia*
[8] *Astrophysics Science Division, NASA Goddard Space Flight Center, 8800 Greenbelt Rd, Greenbelt, MD 20771, USA*
[9] *Cosmic Dawn Center (DAWN), Niels Bohr Institute, University of Copenhagen, Jagtvej 128, København N, DK-2200, Denmark*
[10] *Department of Physics and Astronomy, York University, Toronto, Ontario, Canada*
[11] *Gemini Observatory, NSF's NOIRLab, 950 N Cherry Ave, Tucson AZ, USA.*
[12] *Universidad Nacional de Córdoba, Laprida 854, Córdoba, CPA: X5000BGR, Argentina.*
[13] *Department of Physics and Astronomy, Tufts University, Medford, MA, USA*
[14] *School of Physics, University of New South Wales, Kensington, Australia*
[15] *ARC Centre for Excellence in All-Sky Astrophysics in 3D*
[16] *Department of Physics and Astronomy and PITT PACC, University of Pittsburgh, Pittsburgh, PA 15260, USA*
[17] *University of Pittsburgh, Department of Physics and Astronomy, 100 Allen Hall, 3941 O'Hara St, Pittsburgh PA 15260, USA*
[18] *Department of Statistical Sciences, University of Toronto, 9th Floor, Ontario Power Building, 700 University Ave, Toronto, ON M5G 1Z5, Canada*
[19] *Data Sciences Institute, University of Toronto, 17th Floor, Ontario Power Building, 700 University Ave, Toronto, ON M5G 1Z5, Canada*

## ABSTRACT

The measured ages of massive, quiescent galaxies at $z \sim 3-4$ imply that massive galaxies quench as early as $z \sim 6$. While the number of spectroscopic confirmations of quiescent galaxies at $z < 3$ has increased over the years, there are only a handful at $z > 3.5$. We report spectroscopic redshifts of one secure ($z = 3.757$) and two tentative ($z = 3.336$, $z = 4.673$) massive ($\log(M_*/M_\odot) > 10.3$) quiescent galaxies with 11 hours of Keck/MOSFIRE $K$-band observations. Our candidates were selected from the FENIKS survey, which uses deep Gemini/*Flamingos-2* $K_b K_r$ imaging optimized for increased sensitivity to the characteristic red colors of galaxies at $z > 3$ with strong Balmer/4000 Å breaks. The rest-frame $UVJ$ and $(ugi)_s$ colors of 3/4 quiescent candidates are consistent with $1 - 2$ Gyr old stellar populations. This places these galaxies as the oldest objects at these redshifts, and challenges the notion that quiescent galaxies at $z > 3$ are all recently-quenched "post-starburst" galaxies. Our spectroscopy shows that the other quiescent-galaxy candidate is a broad-line AGN ($z = 3.594$) with strong, redshifted H$\beta$+[O III] emission with a velocity offset > 1000 km/s, indicative of a powerful outflow. The star–formation history of our highest redshift candidate suggests that its progenitor was already in place by $z \sim 7 - 11$, reaching $\sim 10^{11} M_\odot$ by $z \simeq 8$. These observations reveal the limit of what is possible with deep near-infrared photometry and targeted spectroscopy from the ground and demonstrate that secure spectroscopic confirmation of quiescent galaxies at $z > 4$ is feasible only with *JWST*.

Corresponding author: Jacqueline Antwi-Danso
jadanso@tamu.edu

## 1. INTRODUCTION

Understanding the physical processes governing the formation and ultimate quenching of the most massive galaxies in the high redshift Universe is one of the major challenges in modern astrophysics. The existence of massive galaxies at $z > 3$ with stellar masses larger than the present-day Milky Way mass ($M_* > 10^{10.7}$, Papovich et al. 2015) requires extreme physics: intense and rapid star formation followed by abrupt quenching (Glazebrook et al. 2017; Forrest et al. 2020a; Caliendo et al. 2021), leading early massive galaxies to evolve passively with very compact morphologies (Straatman et al. 2014; Wellons et al. 2015; Baggen et al. 2023). Because these galaxies push the limits of astrophysics, they are excellent sites to test galaxy formation models from parsec to gigaparsec scales via e.g., constraining the shape of the initial mass function at early times (Esdaile et al. 2021; Forrest et al. 2022) to tracing the hierarchical growth of dark matter haloes in ΛCDM (Behroozi & Silk 2018).

The myriad physical processes involved in massive galaxy formation, as well as the dynamic range of the spatial and temporal timescales of these processes, beg a plethora of interesting science questions. At the center of these is: *how do massive galaxies in the early Universe assemble their stellar masses so quickly?* Studies addressing this question have made significant progress using with deep near-infrared imaging surveys over the past decade, including, e.g., CANDELS (Koekemoer et al. 2011; Grogin et al. 2011), UltraVISTA (McCracken et al. 2012; Muzzin et al. 2013; Marsan et al. 2022), VIDEO (Jarvis et al. 2013), DES+VHS (Banerji et al. 2015), and COSMOS (Ilbert et al. 2009), pushing the field towards population statistics and detailed characterization of massive galaxies at high redshift.

Despite this progress, detecting these galaxies presents significant observational challenges. Quiescent galaxies have little to no ongoing star formation, hence they lack strong emission features and can only be detected by their stellar continuum. They must be observed in the near-infrared ($1 − 5$ $\mu$m) because they are faint in the rest-frame UV-optical, where the brightest nebular emission lines are. Unfortunately, the near-infrared is prone to numerous systematics, including high backgrounds that vary both spatially and temporally as well as instrumental contamination from thermal sources. These observational challenges have resulted in significant discrepancies in measured galaxy properties (e.g., Alcalde Pampliega et al. 2019) and predictions from cosmological simulations (e.g. Roca-Fàbrega et al. 2021), with number densities differing up to an order of magnitude in certain cases (Merlin et al. 2019, Girelli et al. 2019, Valentino et al. 2023).

The recent discovery of massive galaxies at $z \sim 10$ by *JWST* (Labbé et al. 2023) stands in tension with predictions from ΛCDM (Boylan-Kolchin 2022), and has therefore sparked a new wave of interest and scrutiny on the accuracy of their derived physical properties (e.g., Steinhardt et al. 2022, Endsley et al. 2022, Papovich et al. 2022, van Mierlo et al. 2023). While it is possible that this tension presents an opportunity to reevaluate our understanding of the physics that shapes massive galaxy formation, it is also equally likely that the systematic uncertainties in our observations are severely underestimated. These systematics include issues with photometric selection techniques, effective survey volume, and/or overestimated stellar masses. Therefore, there is a clear and urgent need for large and robust samples of massive galaxies at $z > 4$ with precise redshifts confirmed via spectroscopy. In particular, the spectroscopic confirmation of *already quenched* massive galaxies at these early epochs gives us important constraints on their probable progenitors (e.g. Valentino et al. 2020, Carnall et al. 2023, Nanayakkara et al. 2024), which were very likely the most massive galaxies during the epoch of reionization.

Surveys that include medium bands, such as the Newfirm Medium Band Survey, NMBS (Whitaker et al. 2011) and the FourStar Galaxy Evolution Survey, ZFOURGE (Straatman et al. 2016), which split the $J$ ($\lambda_c = 1.235\mu m$) and $H$ ($\lambda_c = 1.662\mu m$) bands, have been our best attempts at addressing these systematics from the ground by improving photometric redshifts and the resulting physical parameters. These medium-band surveys have been instrumental in the discovery of massive quiescent galaxies up to $z \sim 3.5$ (e.g., Marchesini et al. 2010, Tomczak et al. 2014, Spitler et al. 2014). Recent programs from *JWST* also show much promise by leveraging medium-band photometry (e.g. JEMS, the *JWST* Extragalactic Medium-band Survey; Williams et al. 2023). In general, medium-band surveys increase the detection rate and fidelity of quiescent galaxy selection by providing higher resolution sampling of the Balmer/4000 Å breaks of this population. This results in tightly constrained photometric redshifts and decreases the fraction of star-forming contaminants, whose emission lines can boost broadband fluxes, mimicking a Balmer break. Even with these improvements, the discovery space from the ground has been limited to the brightest quiescent galaxies at $z < 4$, because at $z > 4$, the Balmer/4000 Å break shifts into the $K$-band, where the thermal background is $10 − 12$ mag brighter than the average source in the field.

The latest medium-band survey pushing the frontiers of ground-based NIR observations is the F2 Extragalactic Near-IR K-Split Survey (FENIKS; Esdaile et al. 2021). Similar to its predecessors, FENIKS uses two new custom-built filters installed on Gemini/*Flamingos-2*. The filters split the $K$-band into a bluer, $K_b$ ($\lambda_c$ = 2.0 $\mu m$), and a redder, $K_r$ ($\lambda_c$ = 2.3 $\mu m$) filter (each with $\Delta\lambda$ = 0.26 $\mu m$), for improved identification of galaxies with strong Balmer/4000 Å breaks at 4.2 < $z$ < 5.2. Combined with its custom-built image processing pipeline designed to remove the high levels of sky noise at these wavelengths and account for spatial variations in the point spread function (PSF) of the imaging and large area (0.6 sq degrees when complete) the 170-hour FENIKS survey is poised to detect up to 120 massive quiescent galaxies at 3 < $z$ < 6 (based on extrapolations from existing stellar mass functions), with < 3% photometric redshift uncertainties and a < 5% outlier fraction.

In this *Paper*, we present Keck/MOSFIRE spectroscopy of three faint ($K_s$∼ 23 − 24 AB) massive quiescent galaxy candidates and one AGN candidate at 3 < $z$ < 5 identified in the FENIKS survey. In Section 2, we describe the selection of targets and our Keck/MOSFIRE spectroscopic program targeting these galaxies. In Sections 3 and 4, we describe how we estimate their redshifts, stellar population parameters, rest-frame colors, and star formation histories, taking into account all of the aforementioned systematics. Finally, we discuss the implications of their discovery in the context of high redshift massive galaxy studies in Section 5. Throughout, we assume a $\Lambda$CDM cosmology with $\Omega_M$ = 0.3, $\Omega_\Lambda$ = 0.7 and $H_0$ = 70 km s$^{-1}$ Mpc$^{-1}$. Rest-frame colors are quoted in AB magnitudes (Oke & Gunn 1983). We adopt a Chabrier (2003) initial mass function (IMF) throughout the paper unless explicitly stated.

## 2. OBSERVATIONS

### 2.1. *Photometry and Sample Selection*

For this study, we identify galaxy candidates in the COSMOS field using imaging from the FLAMINGOS-2 Extragalactic Near-Infrared K-band Split (FENIKS) survey (Esdaile et al. 2021). FENIKS is a 170−hr Gemini Large and Long Program (LLP) that uses two novel medium-band filters ($K_b$, $\lambda_c$ = 2.0 $\mu m$ and $K_r$, $\lambda_c$ = 2.3 $\mu m$, with $\Delta\lambda$ = 0.26 $\mu m$) that "split" the $K$-band observing window. As such, FENIKS is a successor to the NMBS (Whitaker et al. 2011) and ZFOURGE (Straatman et al. 2016) surveys, which split the $J$ and $H$ bands for higher resolution sampling of the Balmer/4000 Å breaks of massive galaxies at 1 < $z$ < 3. In a similar fashion, the $K_bK_r$ data are sensitive to the Balmer/4000 Å breaks of massive galaxies at $z$ > 4, which ensures more accurate photometric-redshift uncertainties of <3%, and also more accurate constraints on galaxy stellar masses from fitting their spectral energy distributions (SEDs; Muzzin et al. 2009). We demonstrated these in our pilot survey (Esdaile et al. 2021).

The FENIKS Gemini LLP covers three extragalactic fields (COSMOS, CDFS, and UDS) with ancillary data from ground-based NIR surveys and *HST* (CANDELS/3D-HST) (Grogin et al. 2011), UltraVISTA (McCracken et al. 2012), UKIDSS (Lawrence et al. 2007). When completed, the wide survey area of ≈ 0.6 sq deg will be comparable to that of the largest *JWST* Cycle 1 program (COSMOS-Web; Casey et al. 2023. This reduces cosmic variance by a factor of 3 (Somerville et al. 2004), and makes FENIKS an excellent community resource for selecting followup targets for *JWST*. We describe the FENIKS catalogs and data reduction for the 0.24 sq deg observed to date in an upcoming paper (Antwi-Danso et al., in prep).

We created photometric catalogs by detecting in deep ($K_s$=25.2 AB, 5$\sigma$, $D$ = 2.1″) images from the third data release of the UltraVISTA survey (McCracken et al. 2012; Marsan et al. 2022). The UltraVISTA catalogs include UV-NIR photometry spanning 49 bands. These were supplemented with mid-IR *Spitzer*/MIPS 24 $\mu m$ observations and far-IR observations from *Herschel* 100 $\mu m$ and 160 $\mu m$ from the *Herschel* PACS Evolutionary Probe (PEP; Lutz et al. 2011), and 250 $\mu m$, 350 $\mu m$, and 500 $\mu m$ observations from the *Herschel* Multi-Tiered Extragalactic Survey (HerMES; Oliver et al. 2012). Objects with a ≥ 3$\sigma$ detection in any of the *Herschel* bands were matched to UltraVISTA sources within $r$ = 1.5″ and stellar population parameters derived with `MAGPHYS` (da Cunha et al. 2008). These catalogs are described in detail in (Martis et al. 2016, 2019).

One out of our four objects (COS55-126891) has a *Herschel* counterpart. For unmatched sources (COS55-128636), upper limits were placed on the far-IR fluxes using the depths of the PEP and HERMES surveys (3$\sigma$ upper limits of 4.5, 9.8, 9.5, 8.1, and 11.4 mJy at 100, 160, 250, 350, and 500 $\mu m$, respectively. The remaining two objects were not included in the modeling from Martis et al. (2016, 2019) as they are fainter than the $K_s$ magnitude cut used. The inclusion of the limits from the far–IR data from *Herschel* ensures stronger constraints on the estimated SFRs and stellar masses compared to those estimated from modeling the UV-to-NIR photometry alone (discussed further in Section 5).

The F2 $K_b$ and $K_r$ images have a native "seeing" (PSF FWHM) ≃ 0.5″. We matched these to the PSF of the UltraVISTA images ($K_s$ FWHM ≃ 1.05″) with an accuracy of 1.5% for aperture diameters, $D$, larger than $D$ > 0.83″. Our $K_b$ and $K_r$ fluxes were measured in optimal $D$ = 0.83″ circular apertures using SEP (Barbary 2016), a Python-based package containing all the core libraries of Source Extractor (Bertin & Arnouts 1996) (traditionally used for source detection and image analysis). The $K_b$ and $K_r$ depths in these optimal apertures are 23.6 AB and 23.02 AB, respectively. We provide more details on our catalogs in the FENIKS pilot survey paper (Esdaile et al. 2021) and our upcoming LLP paper (Antwi-Danso et al. in prep).

Our spectroscopic targets were selected from one of our 6.2′ diameter pointings in COSMOS ($\alpha$ : $10^h01^m49.1016^s$, $\delta$ : +02°28′12.08″). We selected galaxies by imposing three criteria: log$M_*/M_\odot$ > $10^{10}$, $K_s$≤ 24.5 mag, and $|K_b - K_r|$ > 1 mag. The latter is indicative of a Balmer/4000 Å break, which falls between these bands at $z$ = 4.2 − 5.2, indicating high mass-to-light ($M/L_V$) ratios. We also selected galaxies at $z$ = 3 − 4 with two color selection techniques, the traditional UVJ diagram (e.g., Williams et al. 2009) and the $(ugi)_s$ diagram (Antwi-Danso et al. 2023). They were selected as galaxies with red $U - V$ ($u_s - g_s$) blue $V - J$ ($g_s - i_s$) colors (Figure 2), bringing our initial sample selection to 17.

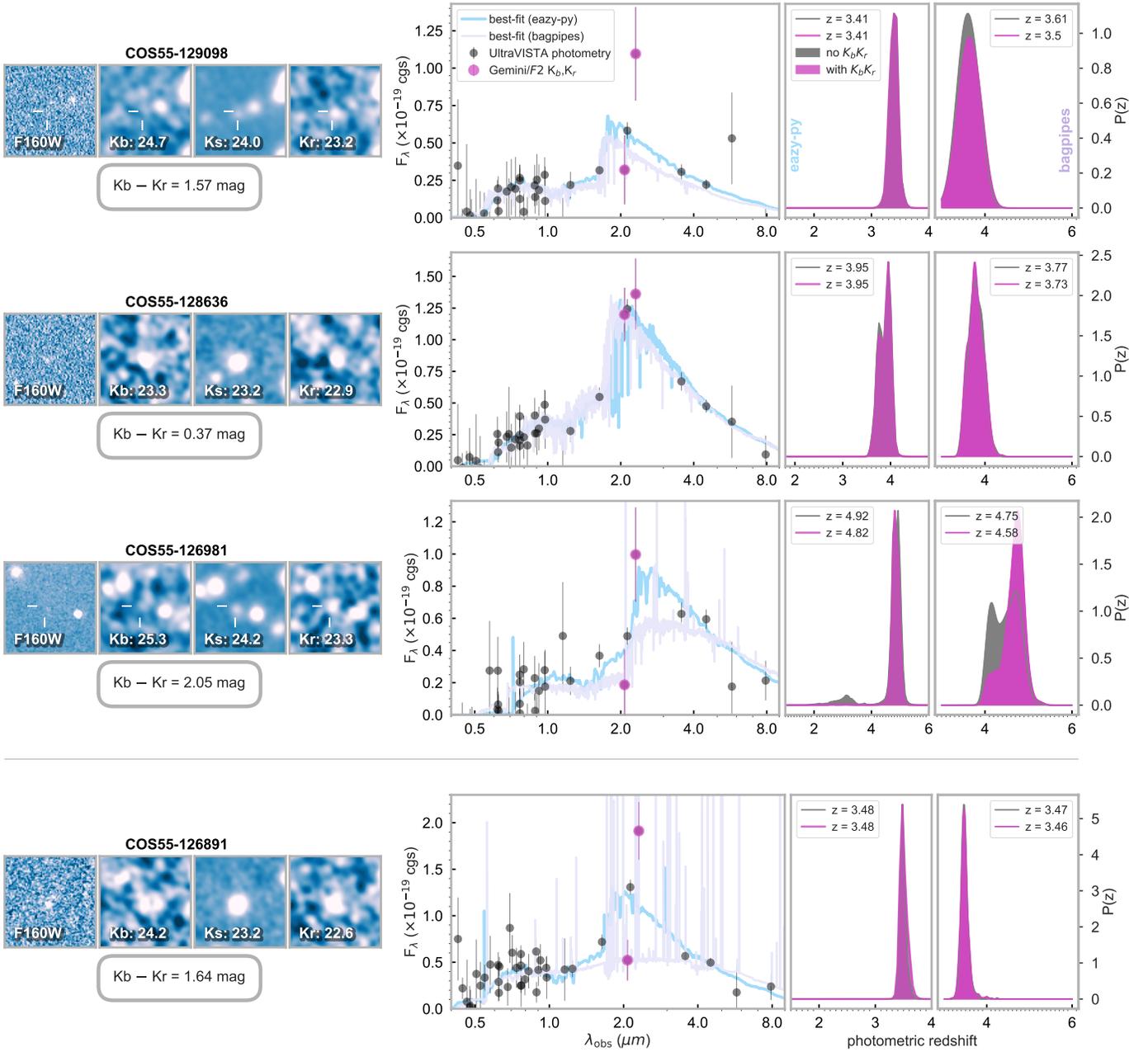

**Figure 1.** 11″× 11″ image stamps aligned north-east and photometric redshift solutions for our four $\log(M_*/M_\odot) > 10$ quiescent candidates and broad-line AGN. The $K$-band images are convolved to match a PSF with FWHM=1.05″. The faintest candidates have markers next to them to distinguish them from neighboring galaxies. Our candidates are either undetected or faint in *HST*/F160W (corresponding to the *H*-band and displayed in the native resolution), and brightest in $K_r$ imaging from Gemini/Flamingos-2. $|K_b - K_r| > 1$ identifies galaxies with strong Balmer/4000 Å breaks (e.g., COS55-128636) and those with strong emission lines (e.g., COS55-126891). In particular, the latter case turns out to be a broad-line AGN with very strong H$\beta$+[O III] emission as indicated by its red $|K_b - K_r|$ color. Our photometric redshift solutions from `eazy-py` and Bagpipes with (magenta) and without (gray) the $K_bK_r$ data are consistent with each other. These $P(z)$ distributions are based on fits to the photometry only. Although the redshift is allowed to vary from $z = 0 - 6$ in both our `eazy-py` and Bagpipes fits, we truncate the displayed range to show all pertinent features of the $P(z)$. The inclusion of the $K_bK_r$ data rules out lower redshift solutions for all candidates but COS55-129098, which has a dusty star-forming solution at $z = 2.77$.

From this, we visually inspected their best-fit SEDs and eliminated candidates with poorly-constrained $P(z)$ distributions (i.e. multimodal solutions with > 2 peaks) and objects that were not detected at S/N > 2 in either $K_b$ or $K_r$, bringing our final selection to a sample of 4 candidates.

In Figure 1, we show the PSF-matched images of these candidates from the Gemini/F2 $K_bK_r$ and UltraVISTA $K_s$ imaging. We also show F160W imaging from the COSMOS-DASH survey (Mowla et al. 2019) in its native resolution. Our quiescent candidates are either faint ($H > 24.4$ mag) or undetected in F160W, which corresponds to a rest-frame of ∼3000 Å at $z = 4$ (i.e., the rest-frame $U$-band). This is indicative of the presence of red, rest–frame UV-optical colors, which we interpret as strong Balmer/4000 Å breaks in these galaxies. The $K_bK_r$ data are important because the *Spitzer*/IRAC data could be contaminated by light from neighboring stars or galaxies (Labbé et al. 2013), particularly for our highest-redshift candidates (e.g., COS55-126981). Due to the relatively narrow widths of these filters, large $|K_b - K_r|$ colors could also correspond to high [O III] + H$\beta$ equivalent widths (rest-frame, 300 - 900 Å) at $z = 3.5$. Hence, the $K_bK_r$ filters are also able to identify galaxies with strong emission lines (e.g., COS55-126891) that may otherwise have colors of quiescent galaxies (see, e.g., the discussion in Antwi-Danso et al. 2023). For these reasons, while the best-fit redshifts of the galaxies in Figure 1 do not change with the inclusion of the $K_bK_r$ data, the filters are important for confirming the strength of the break in COS55-128636 and COS55-126981 (i.e., it is unlikely that the break is mimicked by emission lines contaminating all three $K$−band filters). Consequently, the filters reduce the likelihood of lower redshift solutions, thereby increasing our confidence in our galaxies' high-redshift solutions.

## 2.2. *Keck/MOSFIRE Observations*

We observed our candidates with MOSFIRE (McLean et al. 2012), a multi-object near-infrared spectrograph installed on the Keck I telescope on the Mauna Kea mountain in Hawaii. With its 6′×3′field of view, it can simultaneously observe up to 46 slits per mask, with a slit width of 0.7″and a spectral resolution of $R \sim 3620$ in the $K$-band. Our observations (PI: Nanayakkara) targeted quiescent-galaxy candidates that fall within a single MOSFIRE field of view. The observations took place over two half nights during March 12 – 13, 2022, and had a median seeing of 0.6″. Our primary targets are four quiescent candidates at $z > 3.5$ (Figures 1 and 3) and our secondary targets included 6 massive star-forming galaxies and 23 extreme emission line galaxies at similar redshifts as filler targets. We will present the observations for the latter samples in a forthcoming paper.

In general, massive quiescent galaxies have received limited spectroscopic coverage (there are less than 60 spectroscopically-confirmed quiescent galaxies at $z > 3$ (Figure 4). Lower-mass, bluer galaxies are often prioritized in spectroscopic surveys because they have prominent emission lines, hence require much less observing time. Although they are located in the COSMOS field, which has been thoroughly mined by the largest imaging surveys e.g., COSMOS-Web (Casey et al. 2022), COSMOS (Weaver et al. 2022), Ultra-VISTA (McCracken et al. 2012), none of our targets were observed by any known spectroscopic surveys of which we are aware, e.g., VANDELS (Pentericci et al. 2018), MOSDEF (Kriek et al. 2015), VUDS (Le Fèvre et al. 2015).

We followed a similar observing procedure with MOSFIRE as implemented by other similar spectroscopic campaigns targeting quiescent galaxies at $z > 3$ (Schreiber et al. 2018, Forrest et al. 2020b). We observed our targets with five masks in the $K$-grating (1.95 − 2.4 $\mu$m). This covers the spectroscopic features of interest, including the redshifted Balmer/metal absorption lines of $z$>3.3 galaxies and [O II] and H$\beta$+[O III] emission lines for galaxies at $3 \lesssim z \lesssim 4$ (e.g., Schreiber et al. 2018, Gupta et al. 2020). All masks were observed at the same position angle and contained at least one "slit star" to (i) track the spatial resolution, (ii) check for "slit drift" (Hutchison et al. 2019), and (iii) measure photometric stability (see Appendix A). We observed with an "ABAB" dither pattern, nodding along the slit by ±2.5″ around the target centroid. Individual exposures lasted for 180s because this has been shown to result in the best background subtraction for quiescent galaxies at these redshifts (e.g., Schreiber et al. 2018). Our total on–source integration time was 11 hours. We list the integration times, average seeing, and average airmass per MOSFIRE mask in Table 1.

We reduced the data with the standard MOSFIRE data reduction pipeline (Prochaska et al. 2020) following the procedure in Nanayakkara et al. (2016). The DRP produces background-subtracted, rectified, and flat-fielded 2D spectra and associated variance for each slit within a given mask. We visually inspected the 2D spectra in each mask for evidence of the spectral features of interest. In particular, our inspections showed that COS55-126891 exhibited a strong feature that we associated with broad-H$\beta$+[O III] emission, visible in a single exposure (of only 3 minutes!). The continuum for the other quiescent candidates was visible in at least two masks, which we attribute to the excellent observing conditions and compact nature of these galaxies. Because we are interested in detecting the continuum of these faint ($K_s \approx 23 − 24$ AB) galaxies, we perform additional steps to improve the signal-to-noise of our spectra and correct for telluric absorption. These additional steps are based on the expertise gathered from similar observing programs (Schreiber et al. 2018, Valentino et al. 2020, Forrest et al. 2020b).

Our spectral extraction, flux calibration, and telluric correction are detailed in Appendix A, however we summarize those details here. We extracted the 1D spectra of our candidates using the optimal extraction technique detailed in Horne (1986). We first identified the expected center of the slit (corresponding to the peak of the source) by summing the flux in the spectral direction. For our faintest sources, we also masked prominent skylines to improve the determination of the centroid of the spatial profile. We used an initial extraction box width of 9 pixels and adjusted this box size based on the updated spatial profile after masking skylines (Appendix A).

**Table 1.** Keck/MOSFIRE observation summary.

| Mask | Observing Date | Integration Time ($ks$) | Average Seeing ($''$)[a] | Average Airmass |
|---|---|---|---|---|
| FENIKS_COSMOS55_22A_4 | March 12, 2022 | 10.8 | 0.6 | 1.0 |
| FENIKS_COSMOS55_22A_3 | March 12, 2022 | 7.2 | 0.9 | 1.1 |
| FENIKS_COSMOS55_22A_3 | March 13, 2022 | 3.6 | 0.6 | 1.4 |
| FENIKS_COSMOS55_22A_2 | March 13, 2022 | 7.2 | 0.6 | 1.0 |
| FENIKS_COSMOS55_22A_1_1 | March 13, 2022 | 10.8 | 0.7 | 1.0 |

[a] Average seeing derived from a slit star in the mask.

We then collapsed the 2D spectrum in the spatial direction to create the 1D spectrum and weighted this by the inverse variance and spatial profile. This process down-weights exposures with poorer seeing and improves the S/N of our galaxies by up to a factor of 5 over a boxcar (uniform) extraction (this estimate assumes a boxcar that is wider than the spatial profile). We used slit stars (i.e., stars intentionally targeted on the same MOSFIRE set-up as our galaxy targets) rather than a standard star to perform flux calibration and telluric correction simultaneously on the extracted 1D spectrum from each mask. To create the final flux-calibrated and telluric-corrected spectrum for each target, we excluded 1D spectra from masks where the S/N is ≤ 10% of the coadded spectrum so that the noisiest masks do not reduce the overall S/N of the coadded spectrum. This applies only to the extracted 1D spectrum from the FENIKS_COSMOS55_22A_3 ($t_{exp}$ = 1 hour) mask for COS55-126981. The S/N per pixel for the final, coadded 1D spectra binned to 20 Å are listed in Table 2.

Extracted spectra can be anchored to the photometry in order to account for potential slit losses. We choose not to do this for two reasons: 1) our unresolved sources are well within the Keck/MOSFIRE slits, which means that slit losses are small, ∼ 9%, assuming a 2D Gaussian profile and that the target is centered within 0.07$''$ on the slit[1]; 2) The $K_r$ flux of COS55-129098 is not explained by its spectrum. There is a factor of ∼ 2 discrepancy between the $K_r$ photometry and the continuum flux at those wavelengths (although they agree within the uncertainties). We discuss this at length in Section 4). Additionally, there are no detected emission lines, which might have explained the high $K_r$ flux. Even without this slit loss correction, the extracted 1D spectra of our candidates are remarkably consistent with the photometry. In Figure 3, we show the coadded 2D spectra (also weighted by inverse variance), extracted 1D spectra binned to 20 Å and 70 Å, and the telluric correction. All further analysis is performed on the 20 Å binned spectra. In the following subsections, we detail the methods for this analysis, i.e. identifying spectral features and determining the spectroscopic redshifts of our sources.

## 3. ANALYSIS

### 3.1. *SED Fitting and Rest-Frame Colors*

#### 3.1.1. *Fitting the photometry with* `eazy-py`

We fit the SEDs of our candidates using two independent codes, `eazy-py` (Brammer 2021) and Bagpipes (Carnall et al. 2019b). `eazy-py` is a Python-based SED-fitting code based on the widely-used `EAZY` (Brammer et al. 2008) photo-$z$ code. `EAZY` was built to handle faint galaxy samples with limited spectroscopic redshifts, as we often have with deep NIR photometric surveys. It fits a non-negative, linear combination of empirically-derived templates (in a user-defined list) to the observed photometry. Two features that distinguish `EAZY` from other photometric redshift fitting codes and make it ideal for fitting high redshift galaxies are (1) a template error function, which seeks to account for wavelength-dependent corrections of the templates, such as variations in the dust extinction law and missing spectral features; and (2) an apparent magnitude prior, which assigns low probabilities to high-redshift solutions for extremely bright galaxies. We use a set of 10 templates which model the following galaxy populations: emission line galaxies, galaxies that that are both old and dusty, old quiescent galaxies, and young, recently-quenched galaxies ("post-starbursts"). We also fit the galaxies with newer templates from FSPS (Conroy & Gunn 2010), `fsps_QSF_12_v3`, and obtain similar results as with the `tweak_UVISTA_v4.1` template set for all except COS55-130302. We discuss this in Section 4.

We also derive rest-frame $U - V$, $V - J$ (Williams et al. 2009) and $u_s$-$g_s$, $g_s$-$i_s$ (Antwi-Danso et al. 2023) colors for our candidates (Figure 2). `eazy-py` determines rest-frame colors by doing a "weighted interpolation." It refits the templates to the data, weighting more strongly the observed photometry that is nearest the rest-frame band (in wavelength) and down-weights photometry that is farther away. The rest-frame colors are then interpolated from the model fluxes flanking the rest-frame band of interest. Rest-frame fluxes derived from the best-fit SED are heavily influenced by the choice of templates and assumed star-formation history, and can vary up to 0.3 mag based on these choices (Merlin et al.

---
[1] A centering offset of > 0.07$''$ increases the slit losses to ∼ 12%

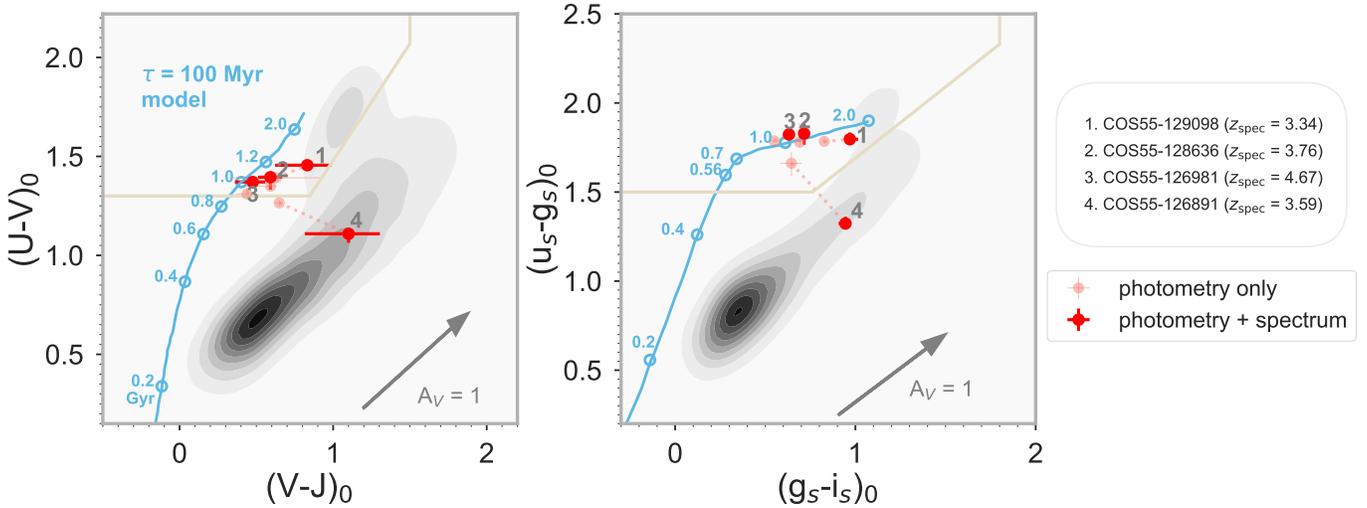

**Figure 2.** Rest-frame *UVJ* and $(ugi)_s$ colors of our candidates derived with photometry alone (small circles) and photometry + spectra (large circles). The color selection lines are calibrated on UltraVISTA data at $1 < z < 2.5$, shown as contours. The gray vectors indicate 1 mag of dust extinction assuming a Calzetti et al. (2000) curve. While the $(ugi)_s$ filters avoid the strongest emission lines, galaxies with strong emission lines ($EW_0 \gtrsim 200$ Å e.g., COS55-126891), can contaminate the quiescent region due to the relatively narrow width of the $g_s$ filter. We also show the color evolution tracks and corresponding ages (in Gyr) of a simple stellar population with an exponentially-declining SFH (*e*-folding timescale $\tau = 100$ Myr). The rest-frame colors of our four quiescent candidates are consistent with $\sim 1 - 2$ Gyr old stellar populations, which challenges the widely-accepted notion that quiescent galaxies at $z > 3$ are all young ( < 300 Myr old) and recently-quenched.

2018). `eazy-py` mitigates these problems by using empirical template sets (which do not assume a star formation history (SFH)) and by using all the available photometry, weighting more strongly bands closest to the rest-frame band of interest. This way, the estimated uncertainties on the rest-frame colors are not prone to the limitations and biases of the chosen template set and SFH. Stellar population parameters from `eazy-py` are computed using a Chabrier IMF (Chabrier 2003).

### 3.1.2. *Jointly fitting photometry and spectra with BAGPIPES*

We also derive photometric and spectroscopic redshift solutions for our candidates using BAGPIPES (Carnall et al. 2018). BAGPIPES is a Bayesian SED fitting code with the functionality to fit both photometry and spectra using on-the-fly model generation and fitting via nested sampling. BAGPIPES has been used by several groups to derive stellar population parameters and investigate the star formation histories of massive galaxies at high redshift (e.g., Carnall et al. 2019b, Shahidi et al. 2020, Wild et al. 2020, Zhuang et al. 2022). We adopt the same nine-parameter model in Carnall et al. 2020 and Carnall et al. 2022. See Table 1 in both papers for a full list of our free parameters and their corresponding priors and a more detailed description of these choices. Stellar population parameters computed using BAGPIPES assume a Kroupa & Boily 2002 IMF. Stellar masses derived using this IMF vary by 0.05 dex from those derived using Chabrier (2003) (Bernardi et al. 2018).

In summary, we allow the redshift to vary from $z = 0 - 6$ with a uniform prior and adopt a double-power-law SFH, which has been shown to reproduce the rising star formation histories of massive galaxies at high redshift (Lee et al. 2010, Papovich et al. 2011, Reddy et al. 2012, Carnall et al. 2019b). The falling and rising slopes ($\alpha$ and $\beta$) are each allowed to vary from $0.01 - 1000$ with a logarithmic prior. The total stellar mass formed by the observed redshift of each galaxy is modeled with a uniform prior from $0 < \log M_*/M_\odot < 13$. It should be noted here that the total stellar mass in BAGPIPES is determined by integrating the star formation history, as opposed to the "living stellar mass," which accounts for the mass in living stars but excludes stellar remnants. The living stellar mass is typically 0.25 dex less than the total stellar mass, although this depends on the adopted star formation history and IMF (Shimizu & Inoue 2013, Carnall et al. 2018).

The stellar and gas phase metallicity is allowed to vary from $0.2 - 0.5$ $Z_\odot$ with a logarithmic prior. This range is consistent with abundance measurements of quiescent galaxies at $z > 1$ (e.g. Kriek et al. 2019, Carnall et al. 2022). Dust attenuation is modeled using Salim et al. (2018), which includes a power law deviation, $\delta$, from the Calzetti et al. (2000) dust curve. $\delta$ is allowed to vary within ±0.3 with a Gaussian prior (with mean, $\mu = 0$, and standard deviation, $\sigma = 0.1$). The V-band attenuation ($A_V$) and strength of the 2175 Å bump (B) are allowed to vary from $0 - 8$ mag and $0 - 5$, respectively, each with a uniform prior. Emission lines are included in the fit using the latest version of the CLOUDY photoionization code (Ferland et al. 2017), with the ionizing parameter (log U) allowed to vary from −4 to −2 with a uniform prior.

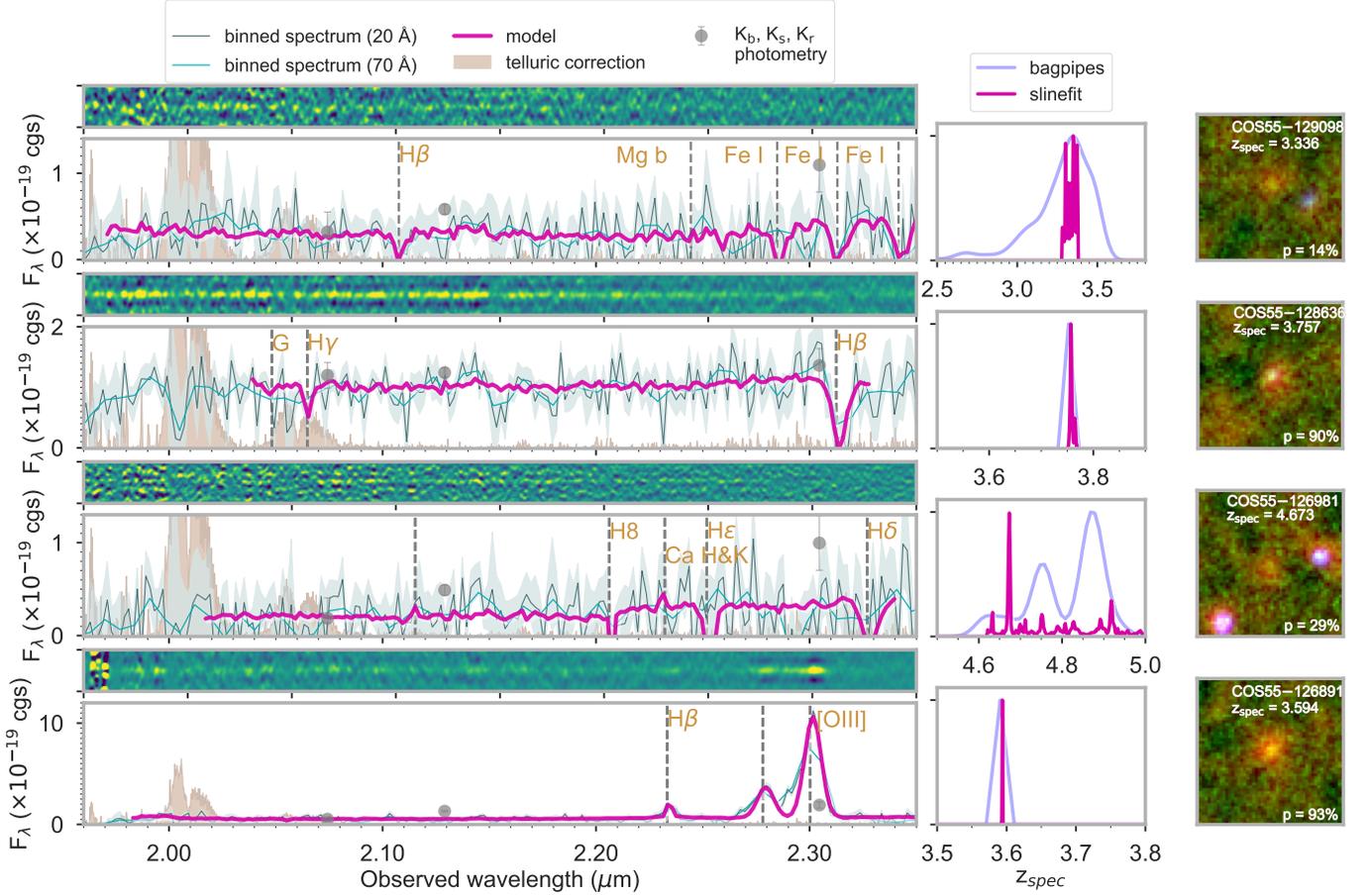

**Figure 3.** From left to right: 11-hour MOSFIRE spectra, redshift probability distribution, and false-color images of our candidates arranged in the same order as Figure 1. For display purposes, we smoothed the 2D spectra with a 0.7″ FWHM Gaussian in the spatial direction. Note that the displayed 2D spectra are the *original*, not telluric-corrected, 2D spectra. We also show the telluric-corrected 1D spectra weighted by the continuum and boxcar smoothed to 20 Å and 70 Å to enhance visibility of the detected absorption lines. We show the best-fit model and *P(z)* from `slinefit` (spectrum only) and the *P(z)* from BAGPIPES (spectrum and photometry). We also show the $K_b$, $K_s$, and $K_r$ points to highlight how consistent the photometry and spectra are, even though they were flux calibrated independently. The false-color images are composed of $K_r$ (red channel), $K_bK_r$ average (green), and $K_b$ (blue), all with the original Gemini/*Flamingos-2* resolution (PSF FWHM ≈ 0.5″) in linear scaling. For reference, a source that is flat in $F_\nu$ would be white in these false-color images. Our candidates are red, which is consistent with the presence of older stellar populations or emission lines. *p* is the probability that $z_{spec}$ lies within ±0.01 of the best-fit redshift (Equation 2).

In general, parametric models impose strong priors on derived stellar masses, star formation rates, and ages (Carnall et al. 2019a). This becomes particularly important when venturing into relatively new and unconstrained parameter space, i.e. the star formation histories and ages of the earliest quiescent galaxies. Others have mitigated this using complex (multiparameter) star formation histories (e.g., Schreiber et al. 2018) and "nonparametric" models (e.g., Leja et al. 2019a, Iyer et al. 2019). The choice of parameters detailed above makes the fitted model flexible enough to permit all potential physically-allowable solutions over the range of specified observed redshifts. We also mitigate this problem by jointly fitting the 51-band UV - NIR photometry with the Keck/MOSFIRE spectra, as fitting spectra and photometry simultaneously has been shown to reduce the impact of degeneracies on derived parameters (Leja et al. 2017, D'Eugenio et al. 2021).

For the spectral fitting, the velocity dispersion is determined by convolving the fit to the spectra with a Gaussian kernel in velocity space, with $\mu$ allowed to vary from 0 − 500 km s$^{-1}$ with a logarithmic prior. Finally, we fit two second-order multiplicative Chebychev polynomials to the spectra to account for imperfections in the flux calibration of the spectra. We mask the edges of each spectrum (first and last 200 Å) because the S/N drops rapidly in those regions. In Figure 5, we compare the photometric redshifts derived using `eazy-py` and BAGPIPES (fit only to the photometry) to the spectroscopic redshifts derived using `slinefit` (de-scribed in Section 3.1.3 below). In Table 2 and Figure 6, we show the star formation histories and recovered parameters for each galaxy, $t_{quench}$, which BAGPIPES determines as the time at which its sSFR fell below 0.2 divided by the Hubble time (e.g., Pacifici et al. 2016), and $t_{50}$, the time at which the galaxy formed half of its stellar mass.

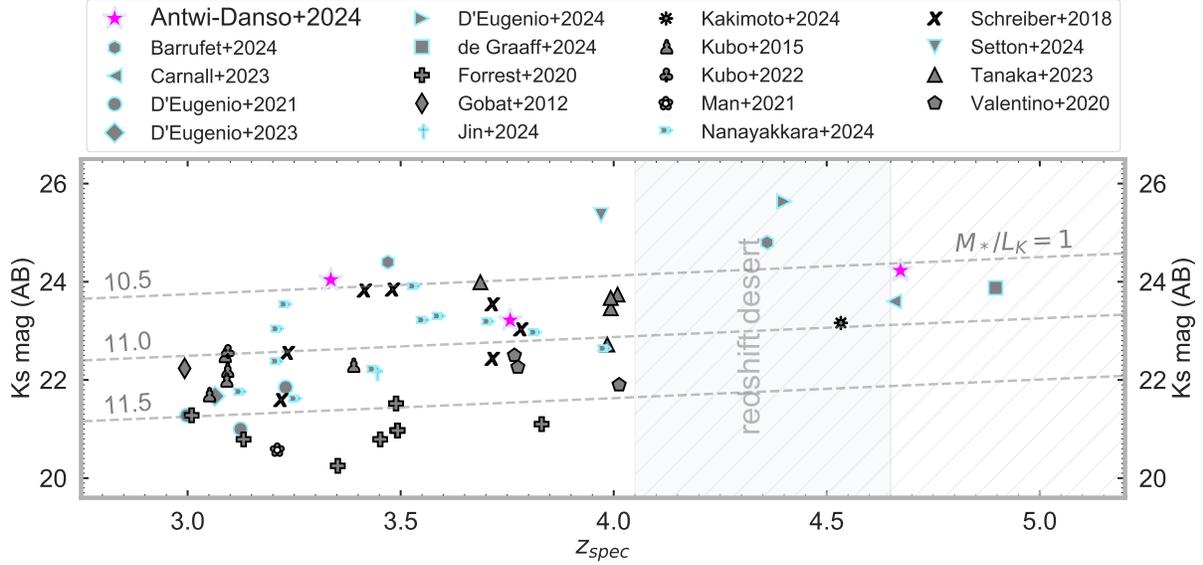

**Figure 4.** $K_s$ magnitude of all 58 spectroscopically-confirmed massive quiescent galaxies at $z \geq 3$. The dashed lines are adapted from Schreiber et al. (2018) and show the $K_s$ magnitude corresponding to $\log(M_*/M_\odot) = 10.5, 11$, and $11.5$, assuming $M_*/L_K = M_\odot/L_\odot$. Cyan points are galaxies observed with *JWST* or *HST* (D'Eugenio et al. 2021). The FENIKS galaxies are some of the faintest quiescent galaxies discovered at these redshifts. These observations reveal a "redshift desert," a glaring dearth of ground-based spectroscopy at $z > 4$. This is partly due to fewer robust photometric candidates at these redshifts, resulting from a lack of bandpasses between $K_s$ and IRAC ch1, which would sample the Balmer break at $z \sim 4 - 5$ (see Section 5 and Figure 7). This bandpass issue is mitigated by *JWST*, as evidenced by the growing number of spectroscopic confirmations at $z > 4$.

### 3.1.3. *Fitting Spectra with Slinefit*

We also estimate redshifts using `slinefit` [2], a publicly-available, spectral fitting code that was developed to fit faint continuum spectra for absorption-line galaxies at high redshift. It has been used by similar studies (Glazebrook et al. 2017, Schreiber et al. 2018, Forrest et al. 2020b, Valentino et al. 2020). `slinefit` performs the fit by doing a $\chi^2$−square minimization between the spectrum and a user-supplied list of templates. For our analysis, we use the same templates as we did with `eazy-py` described in subsection 3.1.1 above. Similar to the fitting with BAGPIPES, we mask the edges of each spectrum. We first explore the fit over a wide redshift range, $0 < z < 6$, in steps of $\Delta z = 0.0003$ and then refine it within $\pm 0.5$ from the initial best-fit redshift using a smaller step size. We also specify the appropriate line ratios for each detected doublet. Without this specification, `slinefit` has no way of knowing a priori what the strongest lines are. The Ca H&K line doublet is fit with a fixed flux ratio of 2 : 3. Similarly, the [O III] doublet was fit with a line ratio of 1 : 3.

---

[2] https://github.com/cschreib/slinefit

To understand the impact of the spectral binning and coadding on our derived redshifts, as well as the limitations of `slinefit`, we performed two tests. First, we ran the fits on the coadded 1D spectra and the individual extractions from each mask. These yielded consistent redshifts. Secondly, we ran `slinefit` on both the unbinned and 20 Å binned spectra of our reliably-detected candidates, COS55-128636 and COS55-126891, and confirmed that the binning process has no impact on the derived spectroscopic redshifts.

The reduced $\chi^2$ of our fits are all close to unity ($0.68-2.64$). To obtain more realistic uncertainties on the redshifts and fitted line parameters, we randomly perturb the spectra using Gaussian noise (where the amplitude is determined by the extracted 1D error spectrum), and refit the spectra with `slinefit`. This results in a $\sim 30\%$ decrease in the S/N of detected lines. The uncertainties listed for the `slinefit` redshifts in Table 2 are representative of the distribution of spectroscopic redshifts derived using this method.

`slinefit` calculates redshift probability distributions using the Benítez (2000) prescription:

$$P(z) \propto \exp\left[\frac{-\chi^2(z) - \chi^2_{min}}{2C}\right] \quad (1)$$

where $C$ is a constant empirical rescaling factor that is related to the noise in our spectra. The default, $C = 1$, assumes ideal, i.e., uncorrelated Gaussian noise. To estimate $C$, we followed a similar procedure as that in Schreiber et al. (2018), by comparing simulated spectra generated with ideal, Gaussian noise with those generated with our MOSFIRE noise spectrum. $C$ is decomposed as $C = C_{\text{ideal}} \times C_{\text{noise}}$, where $C_{\text{ideal}}$ is a correction factor that can be expressed as the ratio between the width of the underlying noise distribution in our spectra and that of the ideal noise spectrum; and $C_{\text{noise}}$ is the ratio of the former and the width of the uncertainty spectrum from the MOSFIRE DRP. We find $C_{\text{ideal}} = 1.25$ and $C = 2.19$, which suggests that our uncertainty spectrum is underestimating the true noise by $\sqrt{C_{\text{noise}}} = 32\%$.

We show the redshift probability distributions from jointly fitting the photometry and spectra with BAGPIPES and fitting the spectra only with `slinefit` in Figure 3. With the exception of COS55-126981 ($z = 4.673$), the redshift peaks from these two independent codes agree with each other. The discrepancy for this target is likely because the fits are based on different spectral features (see Section 4 for a more detailed discussion). With such low S/N, it is difficult to distinguish the higher-order Balmer absorption lines from each other.

## 4. RESULTS

From our analysis, we measure one secure redshift (COS55-128636, $z_{spec} = 3.757$) and two tentative redshifts (COS55-129098, $z_{spec} = 3.336$; COS55-126981, $z_{spec} = 4.673$) for our quiescent candidates. We also measure a secure redshift for the broad-line AGN in COS55-126891 at $z_{spec} = 3.594$. Galaxies with "secure" redshifts meet three conditions: 1) The extracted and binned 1D spectrum used for analysis has a median S/N > 1/pixel, which is required for robust analysis; 2) The best-fit spectroscopic redshifts from `slinefit` and BAGPIPES agree to within 0.5%; and 3) There are no plausible alternative solutions at lower redshifts. To quantify this, we use an adapted version of the robustness criterion, $p$, in Schreiber et al. (2018), defined as

$$p = \int_{-0.01}^{+0.01} \frac{P_1(z) + P_2(z)}{\sqrt{\sigma_1^2 + \sigma_2^2}} dz \quad (2)$$

where $P_N(z)$ and $\sigma_N(z)$ are the redshift probability distribution and associated standard deviation from each fitting code. Secure redshifts are therefore those with $p \geq 90\%$. The estimated $p$ for each galaxy is shown in Figure 3. We provide the photometric and spectroscopic redshifts and physical properties of each galaxy in Table 2. We present their SEDs and magnitudes in Figure 1, Keck/MOSFIRE spectra in Figure 3, rest-frame colors in Figure 2, and star formation histories in Figure 6.

Two of our quiescent candidates (one secure: COS55-128636, and one tentative: COS55-126981) display strong Balmer breaks in their SEDs. Their $K_b$ and $K_r$ colors provide very strong constraints on their photometric redshifts ($\sigma_z = \Delta z/(1 + z_{spec}) \lesssim 3\%$, Figure 5). Our interpretation is supported by the presence of Balmer absorption lines and lack of emission lines in their spectra. Additionally, the false-color images of our quiescent candidates (Figure 3) reveal that they are compact and red, which is typical of quiescent galaxies at high redshift (e.g., Glazebrook et al. 2017, Schreiber et al. 2018). Their blue and red slopes are also well-constrained by the 51-band photometry from UltraVISTA and FENIKS. In all cases, dusty star forming solutions at lower redshift are strongly disfavored by the inclusion of the $K_b K_r$ data and/or spectra in our fitting and analysis. In this section, we briefly discuss the observed properties of each object (in order of increasing redshift) and explore their implications in Section 5.

### 4.1. *Quiescent Candidate COS55-128636, $z_{spec} = 3.757$*

This is the brightest quiescent candidate in our sample ($K_s = 23.2$ AB). The spectroscopic redshifts derived from both `slinefit` and BAGPIPES are based on the presence of a strong absorption lines at 2.062 $\mu$m and 2.31 $\mu$m, which we identify as H$\gamma$ and H$\beta$ at $z = 3.757$. The spectroscopic redshift coincides with the secondary peak of its photometric redshift solution $P(z)$ (see Figure 3). There are no alternative redshift solutions of any significant likelihood based on either the photometry or spectroscopy.

In this galaxy, we detect H$\beta$ and H$\gamma$, which is blended with the G-band. Generally, H$\beta$ and H$\gamma$ should have similar line strengths for 0.5 – 1 Gyr old stellar populations (González Delgado et al. 1999). We see this in other, similar galaxies at this redshift in the literature (Glazebrook et al. 2017; Valentino et al. 2020) . However, in COS55-128636, H$\beta$ is the stronger line. This could be the result of a telluric over-correction. Our slit stars are fainter than the standard star, which means that the telluric correction is based on lower S/N data. This could impact the S/N of bonafide absorption features in the galaxy's spectrum (see Appendix A and Schreiber et al. 2018 for a more detailed discussion about this topic). We therefore suspect the telluric correction is the likely cause for weaker H$\gamma$ absorption in this galaxy here.

### 4.2. *The broad-line AGN in COS55-126891, $z_{spec} = 3.594$*

Cosmological simulations typically invoke AGN feedback to explain the observed properties of massive galaxies. At $z \sim 0$, the galaxies produced by simulations in the absence of AGN feedback tend to be too massive, too compact, too blue, and too bright compared to their observed counterparts (Harrison et al. 2012). With AGN occupying at least 60% of massive galaxies at $z > 3$ (Marsan et al. 2017), there is a growing body of direct observational evidence that places AGN feedback as the primary mechanism for quenching at these redshifts (Schreiber et al. 2018; Kubo et al. 2022; Kocevski et al. 2023; D'Eugenio et al. 2023).

While we initially selected COS55-126891 as a quiescent-galaxy candidate, the MOSFIRE spectrum of COS55-126891 reveals strong, unambiguous H$\beta$+[O III] emission with $f_{\text{[O III]}\lambda5007}/f_{\text{H}\beta} > 6$ and rest-frame $EW_0$ (H$\beta$+[O III]) = 428.5 Å. Given that this galaxy has a large inferred stellar mass of log $(M/M_\odot) = 10.22^{+0.01}_{-0.12}$ (after correcting for emission lines), this is consistent with a broad-line AGN (e.g., Baldwin et al. 1981, Trump et al. 2013, Marsan et al. 2017, Strom et al. 2017). Additionally, this source has an X-ray counterpart in the Marchesi et al. (2016) catalogs and its H$\beta$+[O III] emission is redshifted by $\Delta\lambda$ = 0.32$\mu$m with respect to the systemic wavelength, corresponding to an outflow velocity of ∼ 1300 km/s. This is highly suggestive of an ionized gas outflow powered by an AGN, as has been observed in other massive galaxies at high redshift (e.g., Spilker et al. 2022).

To determine if COS55-126891 is star-forming or quiescent, we would need to disentangle the host galaxy properties from those of the AGN by fitting both components simultaneously. Galaxies with strong emission lines are known to contaminate quiescent samples at these high redshifts and stellar masses (Marsan et al. 2017, Forrest et al. 2020b, Schreiber et al. 2018), highlighting the importance of spectroscopic followup. By mimicking a Balmer/4000 Å break, emission lines can also lead to overestimated stellar masses. When we correct for the presence of emission lines by subtracting the line flux from the photometry and then refitting, the stellar mass of this object decreases by log $M_*/M_\odot$ = 0.67 dex.

The catalog fluxes of this object are below the survey depths of the PEP and HERMES surveys (Section 2.1). Hence, at 350 $\mu$m [3], its 3$\sigma$ flux upper limit is 1.4 mJy, which suggests that the line-of-sight extinction toward the AGN is not significant. This implies that any dust present is not associated with the AGN, negating the need to account for emission from a dusty torus. Although the galaxy is also not detected at 24 $\mu$m, this does not rule out the possibility of dust-obscured star-formation in the host galaxy as even the brightest dusty star-forming galaxies at $z > 3$ are undetected in MIPS (e.g., Lestrade et al. 2010, Casey et al. 2019). Using the Herschel observations (Section 2.1), Martis et al. (2019) derive a dust luminosity, $L_{dust}$ = $10^{10.82} L_\odot$, which corresponds to SFR$_{IR}$ < 6.6 M$_\odot$ yr$^{-1}$ (using Equation 1 in Daddi et al. 2007, where we converted from Salpeter to Chabrier stellar masses by multiplying by a factor of 0.54 e.g., Papovich et al. 2011). In summary, the best-fit model for this object and derived physical properties from BAGPIPES are thus not exhaustive. We defer the two-component SED fitting and detailed study of the outflow kinematics to future work.

### 4.3. *Quiescent Candidate COS55-129098, $z_{spec}$ = 3.336*

This is the lowest redshift quiescent candidate in our sample and the second faintest ($K_s$ = 24.0 AB). With a stellar mass of log $M_*/M_\odot$ = 10.3, it is also one of two *quiescent* galaxies at $z > 3$ with log $M_*/M_\odot$ < 10.5 for which there is spectroscopy. Its rest-frame *UVJ* and $(ugi)_s$ colors are consistent with a stellar population with an age of ∼ 1.5 − 2 Gyr. This is *older* than the other galaxies in our sample, and could be evidence for galactic "downsizing" (Thomas et al. 2005), where the more massive the galaxy, the earlier its formation epoch and the faster its formation time.

While its extracted 1D spectrum is low S/N (median is 0.9/pixel, maximum is 4.2/pixel), this is sufficient to allow us to accurately determine the spatial distribution of the source light (see Figure 9) and hence derive a spectroscopic redshift using stellar absorption features. If this galaxy is quiescent, the most plausible feature (based on its telluric correction) lies at 2.1 − 2.2 $\mu$m in the observed frame. This is best fit by H$\beta$ at $z_{spec}$ = 3.336. The feature at 2.32 $\mu$m is then best fit by the Fe I triplet at 5270 − 5401 Å (although their detection is highly unlikely given the absence of Mg *b*, which tends to be the stronger line in these stellar populations). The secondary peak at $z_{spec}$ = 3.43 favors a detection of Mg *b* at 2.28 $\mu$m. This is closer to its photometric redshift ($z_{phot}$ = 3.41). Given how broad the $P(z)$ from BAGPIPES based on jointly fitting its photometry and spectrum, we cannot rule out either possibility. We therefore report these solutions and maintain a "tentative" redshift for this galaxy.

Finally, its $K_r$ flux is higher than expected given its best-fit SED (2.2× higher at SNR = 3.5 than the model photometry). One possibility is that its $K_r$ flux is contaminated by strong H$\alpha$ emission. This elevated $K_r$ flux due to H$\alpha$ is consistent with a dusty star-forming solution at $z_{spec}$ = 2.77. We disfavor this possibility as there is a lack of emission lines in the observed Keck/MOSFIRE spectrum, although this could also be due to extreme dust obscuration (A$_V$ > 5 mag assuming Calzetti et al. 2000).

### 4.4. *Quiescent Candidate COS55-126981, $z_{spec}$ = 4.673*

The `slinefit` spectroscopic redshift of this galaxy is based on the presence of H$\delta$, H$\epsilon$ (blended with Ca H&K), and H$\zeta$, observed at SNR ∼ 2−3. The BAGPIPES spectroscopic redshift allows a larger range of solutions, primarily because the SNR of the spectroscopic data is relatively low. The peak of the $P(z)$ from the BAGPIPES fit coincides with $z_{spec}$ = 4.862, which is closer to the galaxy's photometric redshift.

BAGPIPES favors a dusty post-starburst solution where the elevated $K_r$ flux is driven by [O II]$\lambda\lambda$ 3727 Å emission at 2.2 $\mu$m. Although there is no evidence for this in the galaxy's spectrum (Figure 3), high dust obscuration (A$_V$ ∼ 7) could be suppressing any nebular emission lines present (e.g., McKinney et al. 2023) or these emission lines could lie outside the MOSFIRE wavelength coverage, as indicated by spectroscopic programs targeting quiescent galaxies at $z > 4$ with *JWST* (Marsan et al. in prep).

We show the star formation histories corresponding to both redshifts in Figure 6. The total stellar mass formed for each SFH (corresponding to the two redshift solutions) only changes by log $M_*/M_\odot$ = 0.07. In Table 2, we only show stellar masses and other stellar population parameters corresponding to the lower (`slinefit`) redshift. It is worth noting when comparing derived redshifts of the same galaxy from two different codes that *lower* redshift does not necessarily mean *more likely*. The $P(z)$ from BAGPIPES indicates that the marginal likelihood (Bayesian evidence) for $z$ = 4.862 is higher, and hence this solution is a better fit for the data than $z$ = 4.673.

Although the continuum is weak in the 2D spectrum, we see a clear continuum peak in the spatial profile of the galaxy from four out of five masks (see Appendix A). The only other spectroscopically-confirmed quiescent galaxy at this redshift in the literature is GS-9209 from Carnall et al. (2023). Although it has a similar magnitude, if confirmed, COS55-126981 would have a higher stellar mass and a more extended SFH (implying a higher $z_{form}$, see Section 5). This is particularly interesting because we derive our stellar population parameters with the same configuration for BAGPIPES as Carnall et al. (2023). This suggests that these differences may not be in methodology, but may be indicative of differences in formation mechanisms and quenching pathways in the highest-redshift quiescent galaxies. Confirmation of COS55-126981 via full coverage of all spectral features of interest with *JWST* will allow us to begin to explore the diversity of the earliest quiescent galaxies.

---

[3] For a galaxy at $z$ = 3.6, the dust peak is at 356 $\mu$m, assuming Wien's law, $T_{peak}$ = 2.9 × 10$^3 \mu$m K/$\lambda_{peak}$, and a dust temperature of 37.5K, from the MAGPHYS catalogs of Martis et al. (2019).

| Galaxy | $z_{phot}$[a] | | $z_{spec}$ | | $\log(M_*/M_\odot)$[d] | SFR | $A_V$ | $t_{50}$ | $t_{quench}$ | age |
|---|---|---|---|---|---|---|---|---|---|---|
| | eazy-py | BAGPIPES | slinefit[b] | BAGPIPES[c] | | ($M_\odot$ yr$^{-1}$) | (mag) | (Gyr) | (Gyr) | (Gyr) |
| 129098 | $3.41^{+0.09}_{-0.14}$ | $3.5^{+0.32}_{-1.1}$ | $3.336^{+0.0056}_{-0.0550}$ | $3.323^{+0.114}_{-0.213}$ | $10.29^{+0.21}_{-0.04}$ | $0.01^{+1.63}_{-0.01}$ | $0.76^{+0.76}_{-0.39}$ | $1.49^{+0.15}_{-0.54}$ | $1.72^{+0.06}_{-0.08}$ | $0.59^{+0.34}_{-0.36}$ |
| 128636 | $3.95^{+0.07}_{-0.24}$ | $3.73^{+0.25}_{-0.12}$ | $3.757^{+0.0011}_{-0.0011}$ | $3.787^{+0.185}_{-0.165}$ | $10.86^{+0.09}_{-0.03}$ | $0.0^{+0.09}_{-0.0}$ | $0.29^{+0.37}_{-0.18}$ | $1.30^{+0.12}_{-0.58}$ | $1.32^{+0.08}_{-0.14}$ | $0.41^{+0.49}_{-0.04}$ |
| 126981 | $4.82^{+0.18}_{-0.56}$ | $4.58^{+0.28}_{-0.15}$ | $4.673^{+0.0012}_{-0.0830}$ | $4.862^{+0.016}_{-0.115}$ | $11.35^{+0.03}_{-0.19}$ | $40.74^{+163.96}_{-34.91}$ | $1.18^{+0.9}_{-0.44}$ | $0.62^{+0.20}_{-0.22}$ | $1.16^{+0.06}_{-0.21}$ | $0.29^{+0.54}_{-0.05}$ |
| 126891 | $3.48^{+0.14}_{-0.06}$ | $3.46^{+0.1}_{-0.06}$ | $3.594^{+0.0003}_{-0.0003}$ | $3.467^{+0.086}_{-0.075}$ | $10.22^{+0.01}_{-0.12}$ | $1.23^{+0.86}_{-0.55}$ | $2.46^{+0.11}_{-0.06}$ | $0.90^{+0.38}_{-0.12}$ | $1.53^{+0.17}_{-0.05}$ | $0.80^{+0.20}_{-0.27}$ |

| Galaxy | median SNR | maximum SNR | $\chi^2_{red}$ | G+H$\gamma$ | | H$\beta$ | | [O III]$\lambda$5007 | |
|---|---|---|---|---|---|---|---|---|---|
| | | | | $f_\lambda$ | EW | $f_\lambda$ | EW | $f_\lambda$ | EW |
| 129098 | 0.88 | 4.15 | 0.75 | – | – | 7.73 ± 9.9 | −6.1 ± 7.9 | – | – |
| 128636 | 3.29 | 8.99 | 1.07 | 21.9 ± 17.1 | −5.38 ± 4.19 | 78.98 ± 23.6 | −15.1 ± 4.6 | – | – |
| 126981 | 0.66 | 3.53 | 0.68 | – | – | – | – | – | – |
| 126891 | 1.91 | 33.83 | 2.64 | – | – | 44.3 ± 10.1 | 14.6 ± 3.6 | 1261.26 ± 19.2 | 413.9 ± 23.0 |

**Table 2.** Top: Redshifts and stellar population parameters of our quiescent candidates and broad-line AGN in the same order as Figures 1 and 3. The derived parameters are from jointly fitting the photometry and spectra with BAGPIPES. SFRs are averaged over the last 10 Myr. $t_{50}$ and $t_{quench}$ are the times (counting from the Big Bang) at which the galaxy formed 50% of its stellar mass and dropped its SFR below 10%, respectively. Ages are mass-weighted ages from BAGPIPES, corresponding to $t_{50}$. **Bottom:** Line properties from slinefit. The SNR values reported are for the boxcar 20 Å binned spectra on which we performed all analysis. Fluxes are in units of $10^{-19}$ erg/s/cm$^2$/Å. Equivalent widths are in rest-frame Ångstroms (negative values indicate absorption). Dashes represent cases where the line flux could not be reliably determined.

[a] Redshifts derived from photometry only (including Gemini $K_b K_r$ photometry).

[b] Derived by fitting spectra only.

[c] Derived by fitting photometry and spectra jointly.

[d] Total stellar mass formed from BAGPIPES as opposed to the "living stellar mass" (described in Section 3.1.2).

## 5. DISCUSSION AND SUMMARY

In this *P*aper, we present near-infrared spectroscopy of three massive quiescent galaxies and one broad-line AGN (COS55-126891) at $3 < z < 5$ discovered in the FENIKS survey. The Gemini-F2 $K_b$ and $K_r$ filters used in the FENIKS survey "split" the K-band, yielding higher resolution sampling of the Balmer/4000 Å break. This improves the identification of these massive galaxies and therefore their photometric redshift accuracy ($\sigma_z < 3\%$, Figure 5). We confirm the quiescent nature and redshifts of one galaxy (COS55-128646, $z_{spec} = 3.757$), via the identification of Balmer absorption lines using 11 hours of Keck/MOSFIRE spectroscopy (Figure 3). We report tentative spectroscopic redshifts for the two other quiescent candidates (COS55-129098 at $z_{spec} = 3.336$ and COS55-126981 at $z_{spec} = 4.673$) given their low S/N spectra and hence weaker constraints from modeling. We derived the star formation histories of our quiescent candidates by jointly fitting their photometry and spectra (Figure 6). In this section, we discuss the implications of these results in the context of other spectroscopically-confirmed quiescent galaxies at $z > 3$.

### 5.1. *Tackling the Age Bias in Quiescent Samples*

First, we consider whether all quiescent galaxies at $z > 3$ are "post-starburst." Most of the spectroscopically confirmed quiescent galaxies at $z > 3$ are young (< 600 Myr) and recently-quenched rather than old, long-dead galaxies (Schreiber et al. 2018, D'Eugenio et al. 2021), as their last star-formation episode occurred over the last ∼0.1 Gyr. These galaxies tend to be bright ($K_s < 22$ AB mag) because, for this age, the light around the Balmer Break in these objects is dominated by A-type stars. A few studies (e.g., Glazebrook et al. 2017, Kalita et al. 2021) have hinted that we may be missing a population of older quiescent galaxies (relative to the age of the Universe at these redshifts).

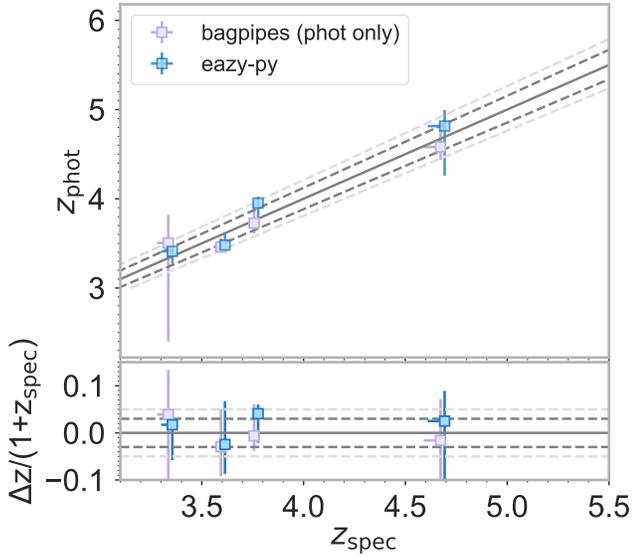
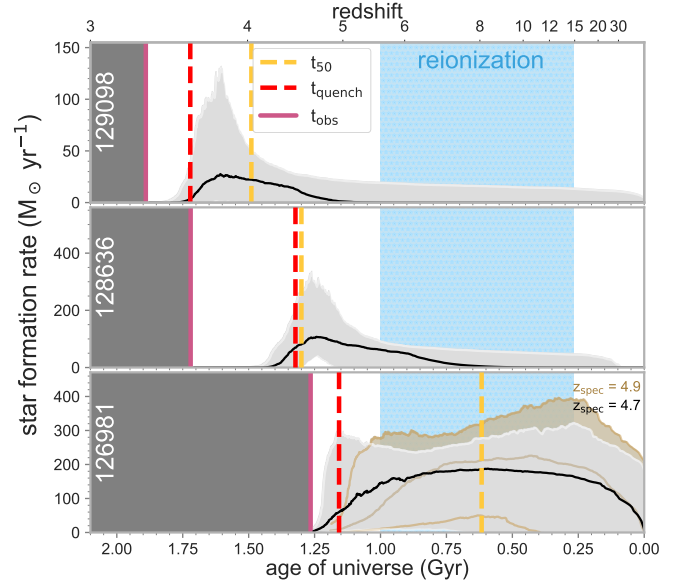

**Figure 5. Top:** Photometric redshifts from `eazy-py` and BAGPIPES vs. spectroscopic redshifts from `slinefit`. The redshifts from BAGPIPES shown here are derived from fitting the photometry alone. We have offset the BAGPIPES points by $z = -0.02$ to make them easier to see. **Bottom:** Photometric redshift scatter ($\sigma_z$). The dark and light gray dotted lines mark the 3% and 5% thresholds. 3/4 of our candidates lie within the former, consistent with predictions from the FENIKS pilot survey. This is a testament to the utility of the $K_b K_r$ filters for selecting robust samples of massive galaxies at these redshifts.

**Figure 6.** Star formation histories of our three quiescent galaxies at $z > 3$, displayed in order of increasing observed redshift. The solid lines show the 50th percentile of the SFH posterior distribution, and the shaded gray regions show the 68% confidence intervals. For each galaxy, we also indicate $t_{50}$ (age at which the galaxy formed 50% of its stars), $t_{\mathrm{quench}}$ (time when the SFR drops below 10% of the past average), and $t_{\mathrm{obs}}$. For COS55-126981, we show the SFHs corresponding to the spectroscopic redshifts from both `slinefit` (black) and BAGPIPES (brown).

These galaxies would have higher $M/L_V$ ratios and be observed $< 500$ Myr after quenching, hence are expected to be $1-3$ mag fainter in the near-IR (Forrest et al. 2020b).

Due to the sensitivity of the deep Gemini $K_b K_r$ imaging to the Balmer break, we have uncovered three quiescent candidates that fit this description. They are the faintest of their kind at these redshifts (Ks ∼ 23 − 24 AB mag), and their rest-frame $UVJ$ and $(ugi)_s$ colors are consistent with *older* stellar populations, with ages of 1–2 Gyr, up to ∼ 90% the age of the Universe at these redshifts. This is consistent with the mass-weighted ages we derived from the star formation histories of these galaxies given the estimated uncertainties (median $1\sigma$ ($3\sigma$) uncertainty of 0.3 Gyr (0.4 Gyr), which is typical for the ages of quiescent galaxies at $z > 3$ e.g., Glazebrook et al. 2017, Carnall et al. 2020). In particular, the SFH models available in BAGPIPES have been shown to produce systematically younger ages, particularly for massive, i.e., log $(M_*/M_\odot) > 10.5$ quiescent galaxies (Carnall et al. 2019a, Kaushal et al. 2023), which suggests that the ages in Table 2 may be underestimated. While the uncertainties on the SFH of COS55-128636 are typical of those galaxies at similar redshifts e.g., Glazebrook et al. (2017), we note that the SFHs of the other two should be interpreted with caution given the lower quality of their spectra and consequently, much more uncertain SFHs.

Additionally, unlike their post-starburst counterparts, our candidates are extremely faint in the rest-frame UV-optical (> 25 AB), which is consistent with their suppressed SFRs and with our photometric selection criteria in Section 2.1. At $z > 3$, the oldest, most quiescent galaxies would push formation times back to the highest redshifts, further straining tensions with theory. For instance, the star formation histories of our $z_{spec} = 4.673$ candidate suggests $z_{form} \sim 8$. This is consistent with recent studies based on *JWST* data suggesting that galaxy formation began earlier than previously thought (e.g., Labbe et al. 2022, Tacchella et al. 2022a, Di Cesare et al. 2023). With such high formation redshifts and old ages, it becomes an even greater challenge for cosmological simulations to both form and quench them at such early times (Boylan-Kolchin 2022). The discovery of these faint, potentially old quiescent galaxies is consistent with recent *JWST* observations confirming that that the prominence of young (< 300 Myr), recently-quenched galaxies at $z > 3$ is simply a selection effect (Glazebrook et al. 2023, Nanayakkara et al. 2024).

## 5.2. *On the Dearth of Confirmed Quiescent Galaxies at $z > 4$*

While it is well established that massive quiescent galaxies exist out to $z \sim 3 - 4$ (Glazebrook et al. 2017, Schreiber et al. 2018, Carnall et al. 2019b, Forrest et al. 2020a, Saracco et al. 2020, McConachie et al. 2022), only a handful have been confirmed with spectroscopy at $z \sim 4$ (Valentino et al. 2020, Nanayakkara et al. 2024, D'Eugenio et al. 2024, Barrufet et al. 2024, de Graaff et al. 2024). Other studies of quiescent galaxies at $z \sim 1 - 3$ have argued their quenching times places them at $z > 4 - 5$ (e.g, Belli et al. 2019; Estrada-Carpenter et al. 2020; Tacchella et al. 2022b).

In spite of this, there is only one quiescent galaxy at $z \sim 4 - 4.6$ that has been confirmed with spectroscopy from the ground (Figure 4, Table 3). This "redshift desert" does not appear to be an observational effect, as H$\gamma$ and H$\delta$ both fall into the MOSFIRE $K$-grating at these redshifts, enabling a robust redshift measurement from the ground. Another possibility is that this is due to a lack of bandpasses between the $K_s$ and IRAC channel 1 bands when observing from the ground. We test this using the best-fit SED of COS-128636, redshifted from $z = 3 - 6$ with each photometric point assigned S/N = 10. We then examine the deviation of the recovered $UVJ$ colors from the true value. As the Balmer/4000Å break moves through the $K_s$ band at $z > 4$, we lose information about the strength of the break until it enters IRAC channel 1 at $z \sim 5.3$. This results in a preference for dusty star-forming solutions at $z \sim 4 - 5$ (Figure 7) and provides an explanation for the missing quiescent galaxies at these redshifts.

This is, of course, a simplified test. In reality, the resulting best-fit solution depends on many factors, such as the number of photometric points, their S/N relative to each other, and template set or assumed star formation history. Testing each of these variables is beyond the scope of this work. Our results however suggest that bonafide quiescent galaxies may be missing from current quiescent samples because they are erroneously fitted with dusty star-forming solutions due to the lack of bandpasses between $K_s$ and IRAC channel 1. Therefore, this is a selection effect that impacts all ground-based photometric searches for quiescent galaxies. This problem can only be resolved from space, due to the morbidly high thermal backgrounds at $\lambda > 2.4$ $\mu$m. Hence, surveys with *JWST* should yield larger samples of quiescent candidates at $z > 4$ due to its continuous wavelength coverage at $1 - 5$ $\mu$m.

In Table 3, we compile all $z > 3$ spectroscopically-confirmed quiescent galaxies with log $(M_*/M_\odot) > 10$ in the literature. There are 58 to-date, with 20 of these being confirmations from *JWST*. With the exception of GS-9209 (Carnall et al. 2023), all the quiescent galaxies $z > 4$ in Figure 4 were selected using *JWST*/NIRCam photometry from either the CEERS (de Graaff et al. 2024 sample) or JADES surveys (Barrufet et al. 2024 sample). This confirms the necessity of bandpasses at $\lambda_{obs} \approx 2 - 3$ $\mu$m for photometric selection of massive quiescent galaxies at $z = 4 - 5$ and validates our prediction for the redshift desert.

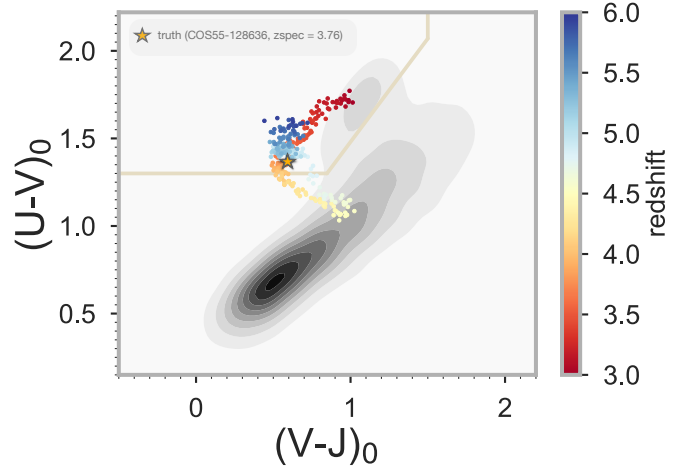

**Figure 7.** *UVJ* colors corresponding to the best-fit SED of COS55-128636 redshifted from $z = 3 - 6$. As the Balmer/4000 Å break moves through the $K_s$ band at $z = 4.2 - 5.3$, we lose information about the strength of the break, resulting in a preference for dusty star-forming solutions. At $z = 5.3 - 6$, we recover that information, as the peak of the Balmer/4000 Å break enters the *Spitzer*/IRAC channels. This suggests that the dearth of quiescent galaxies at $z \sim 4 - 5$ in current catalogs is partly due to the inaccessibility of wavelengths $> 2.4 \mu m$ from the ground. Recent results from *JWST* confirm this.

## 5.3. *Too Big to Be?*

The redshifts and stellar masses of two of our three quiescent candidates imply that they converted their dark matter halo baryons into stars at abnormally high efficiencies ($\epsilon \geq 20\%$), a factor of at least 2 higher than observed in the local Universe (Baldry et al. 2008), assuming a number density of $1.8 \times 10^{-5}$ Mpc$^{-3}$ (Straatman et al. 2014), which corresponds to a dark matter halo mass of $3 \times 10^{12}$ M$_\odot$[4]. Our $z = 4.673$ candidate has a stellar mass of log $M_*/M_\odot = 11.35^{+0.03}_{-0.19}$. At these redshifts, a stellar mass this high requires $\epsilon \sim 98\%$ (assuming a cosmic baryon fraction of 16% from Planck Collaboration et al. 2016), near the maximum rate given predictions from $\Lambda$CDM. The existence of these massive (log $M_*/M_\odot \gtrsim 10.5$), quiescent galaxies is therefore an enigma: a challenge to galaxy formation theory dubbed the *"impossibly early galaxy" problem* (Steinhardt et al. 2016).

Our quiescent candidates have stellar mass estimates from four SED fitting codes which jointly fit the UV-NIR photometry and spectroscopy (BAGPIPES), UV-FIR photometry (MAGPHYS), and UV-NIR photometry (eazy-py and FAST). These stellar masses agree to within $\sim 0.2$ dex with uncertainties on the order of $\sim 0.1$ dex. These are within the expected systematic uncertainties from modeling due to the different assumptions in SED fitting codes (Pacifici et al. 2023). Furthermore, our candidates all have *Spitzer*/IRAC photome-try (3.6 - 8$\mu$m), which improves uncertainties in $M_*$ by a factor of 1.3 dex (Muzzin et al. 2009). Additionally, two of our candidates (COS55-128636 and COS55-126981) have stellar mass estimates computed using FIR data (from the catalogs of Martis et al. 2016, 2019), which improves stellar mass estimates by ∼ 0.1 − 0.3 dex (Leja et al. 2019b).

---
[4] Estimated using HMFcalc (Murray et al. 2013)

The mid- and far-IR observations from *Spitzer* and *Herschel* (see Section 2.1 for details) were used to estimate the stellar masses and stellar population parameters for these two candidates using MAGPHYS in the catalogs from Martis et al. (2016, 2019). Such data is important because it helps discriminate between different sources of dust heating, namely heating due to star formation versus heating from evolved stars (Papovich et al. 2006, Caputi et al. 2012). This ultimately helps constrain the amount of stellar mass that is locked up in old (> 100 Myr) stars. While the stellar mass of the AGN (COS-126891) does not change appreciably ($\Delta M_* = -0.05$ dex), that of the quiescent candidate (COS55-128636) does ($\Delta M_* = +0.17$ dex) due to the inclusion of far-IR limits in the SED fitting. This increase in stellar mass is likely due to differences in the treatment of dust attenuation and emission between FAST and MAGPHYS. The latter includes models that account for emission from polyaromatic hydrocarbons (PAHs), which tend to dominate the mid-IR emission from dust in star-forming regions and therefore can bias stellar mass estimates.

Recent observations targeting massive galaxies suggest that we may be on the cusp of a paradigm shift, specifically that galaxy formation began earlier than previously thought. The launch of *JWST* has proved to be transformative for this science due to the observatory's low sky backgrounds, increased sensitivity, and access to wavelengths > 2$\mu$m. Nanayakkara et al. (2024) obtained low-resolution ($R \sim 50-100$) *JWST*/NIRSpec PRISM observations of 5 unconfirmed candidates at $3 < z < 4$ and 1 confirmed massive quiescent galaxy from the Schreiber et al. (2018) sample of 24 candidates. This gave us, for the first time, a continuous view of the observed $1 - 5\mu$m spectral energy distributions of quiescent galaxies at high redshift. The 5 candidates were confirmed to have the distinctive Balmer break and absorption features typical of post-starbursts. Their spectra also revealed strong H$\beta$+[O III] and H$\alpha$ emission lines, likely powered by AGN.

Additionally, Carnall et al. (2023) confirmed the redshift of a massive quiescent galaxy at $z = 4.658$, when the Universe was just 1.25 Gyr old. The higher resolution ($R \sim 1000$) spectrum of this galaxy allowed for a much more detailed analysis, including estimating its iron abundance and $\alpha$-enhancement, which suggested an extremely short ($\lesssim 200$ Myr) formation timescale. Similar to the Nanayakkara et al. (2024) objects (see also de Graaff et al. 2024), this spectrum of this galaxy also revealed broad H$\alpha$ emission that was significantly higher than expected due to star-formation, but rather, consistent with AGN activity or galactic outflows, which have been observed in post-starbursts at $z > 1$ (Maltby et al. 2019). The derived black hole mass of $\log(M_\cdot/M_\odot) = 8.7 \pm 0.1$, strongly implies the existence of a supermassive black hole which is in line with quenching due to AGN feedback.

The star formation histories of the Nanayakkara et al. (2024) and Carnall et al. (2023) galaxies suggest that their progenitors were already in place by $z \sim 10$, just 500 Myr after the Big Bang. This is consistent with the discovery of compact massive galaxies at $z \sim 7 - 10$ (Labbé et al. 2023, Baggen et al. 2023) and excess of UV-bright galaxies discovered in Cycle 1 surveys (Harikane et al. 2023, Bunker et al. 2023). These galaxies are uncomfortably massive based on model predictions, suggesting variation in the initial mass function (IMF) at higher redshifts (Steinhardt et al. 2016, 2022) or overestimated stellar masses (Endsley et al. 2022, van Mierlo et al. 2023).

While our observations have shed light on these issues, they are far from being completely resolved. Our results have demonstrated that robust spectroscopic confirmation of quiescent galaxies at $z > 4$ is not feasible from the ground, even with deep Keck/MOSFIRE spectra. We need the continuous wavelength coverage from *JWST* to *directly* constrain the ages and formation timescales of massive quiescent galaxies via the detection of age, metallicity, and abundance indicators. The largest factor limiting progress on this front is the lack of high SNR spectroscopic data with continuous wavelength coverage ($1.6 - 5.3 \mu$m) that includes these key features. This is a Herculean task to attempt from the ground, where atmospheric absorption and emission make continuum observations far more challenging, thereby requiring 30+ hours on the most sensitive ground-based spectrographs to make marginally constraining measurements of elemental ratios (e.g., Kriek et al. 2016, 2019, Carnall et al. 2019b, Onodera et al. 2015). With *JWST*, we can now make these measurements to better accuracy. While this is the missing piece in this puzzle, we now have the technological capabilities to resolve this question moving forward.

Table 3. Spectroscopically-Confirmed Massive Quiescent Galaxies at $z \geq 3$

| | Galaxy ID | $\alpha$ | $\delta$ | $K_s$ (AB mag) | $z_{spec}$ | $\log(M_*/M_\odot)$ | References |
|---|---|---|---|---|---|---|---|
| | (1) | (2) | (3) | (4) | (5) | (6) | (7) |
| 1 | COS55-129098 | 150.43732 | +2.463920 | 24.04 ± 0.09 | $3.336^{+0.0056}_{-0.0550}$ | $10.29^{+0.21}_{-0.04}$ | this work |
| 2 | COS55-128636 | 150.45459 | +2.455994 | 23.22 ± 0.04 | $3.757^{+0.0011}_{-0.0011}$ | $10.86^{+0.09}_{-0.03}$ | this work |

Table 3 *continued*

| | Galaxy ID | $\alpha$ | $\delta$ | $K_s$ (AB mag) | $z_{spec}$ | $\log(M_*/M_\odot)$ | References |
|---|---|---|---|---|---|---|---|
| | (1) | (2) | (3) | (4) | (5) | (6) | (7) |
| 3 | COS55-126981 | 150.46120 | +2.429547 | $24.23 \pm 0.11$ | $4.673^{+0.0012}_{-0.0830}$ | $11.35^{+0.03}_{-0.19}$ | this work |
| 4 | 6620 | 53.078727 | −27.839608 | 24.4 | 3.47 | $10.35^{+0.08}_{-0.07}$ | Barrufet et al. 2024, Carnall et al. 2020 |
| 5 | 8290 | 53.081879 | −27.828800 | 24.8 | 4.36 | $10.40^{+0.06}_{-0.06}$ | Barrufet et al. 2024, Carnall et al. 2020 |
| 6 | GOODSS-09209 | 53.108177 | −27.825122 | 23.6 | $4.658^{+0.0002}_{-0.0002}$ | $10.61^{+0.02}_{-0.02}$ | Carnall et al. 2023, Barrufet et al. 2024 |
| 7 | 252568 | 149.452386 | +1.666061 | $21.05 \pm 0.01^\diamond$ | $3.124^{+0.003}_{-0.003}$ | 11.32 | D'Eugenio et al. 2021$\dagger$ |
| 8 | 361413 | 150.504042 | +1.840083 | $21.91 \pm 0.01^\diamond$ | $3.230^{+0.007}_{-0.006}$ | 10.75 | D'Eugenio et al. 2021$\dagger$ |
| 9 | 575436 | 150.182362 | +2.174642 | $21.40 \pm 0.01^\diamond$ | $2.998^{+0.002}_{-0.003}$ | 11.17 | D'Eugenio et al. 2021$\dagger$ |
| 10 | GS-10578 | 53.165778 | −27.814111 | $21.67 \pm 0.02^\spadesuit$ | $3.064^{+0.0004}_{-0.0004}$ | $11.20^{+0.06}_{-0.06}$ | D'Eugenio et al. 2023 |
| 11 | 1080660 | 189.275449 | +62.214135 | $25.63 \pm 0.80^\clubsuit$ | 4.40 | ... | D'Eugenio et al. 2024 |
| 12 | RUBIES-EGS-QG-1 | 214.915546 | +52.949018 | $23.88 \pm 0.17^\spadesuit$ | $4.896^{+0.0006}_{-0.0007}$ | $10.94^{+0.02}_{-0.02}$ | de Graaff et al. 2024 |
| 13 | XMM-VID1-2075 | ... | ... | 20.79 | $3.452^{+0.0014}_{-0.0017}$ | $11.52^{+0.00}_{-0.05}$ | Forrest et al. 2020b |
| 14 | XMM-VID3-1120 | ... | ... | 20.97 | $3.492^{+0.0018}_{-0.0029}$ | $11.47^{+0.02}_{-0.03}$ | Forrest et al. 2020b |
| 15 | COS-DR3-160748 | 150.115875 | +2.563675 | 20.25 | $3.352^{+0.0008}_{-0.0006}$ | $11.46^{+0.01}_{-0.08}$ | Forrest et al. 2020b, Marsan et al. 2015, Saracco et al. 2020 |
| 16 | COS-DR3-201999 | ... | ... | 20.79 | $3.131^{+0.0014}_{-0.0012}$ | $11.40^{+0.03}_{-0.01}$ | Forrest et al. 2020b |
| 17 | XMM-VID3-2457 | ... | ... | 21.52 | $3.489^{+0.0032}_{-0.0024}$ | $11.26^{+0.02}_{-0.03}$ | Forrest et al. 2020b |
| 18 | COS-DR3-84674 | ... | ... | 21.28 | $3.009^{+0.0015}_{-0.0011}$ | $11.25^{+0.03}_{-0.01}$ | Forrest et al. 2020b |
| 19 | COS-DR1-113684 | ... | ... | 21.10 | $3.831^{+0.0014}_{-0.0020}$ | $11.20^{+0.03}_{-0.04}$ | Forrest et al. 2020b |
| 20 | XMM-2599 | 36.792075 | −4.579163 | $20.97 \pm 0.02$ | $3.493^{+0.003}_{-0.008}$ | $11.49^{+0.03}_{-0.02}$ | Forrest et al. 2020a, Shen et al. 2021 |
| 21 | RS-235 | 222.318750 | +8.926306 | $22.24 \pm 0.05^*$ | $2.993^{+0.015}_{-0.015}$ | $11.08^{+0.15}_{-0.12}$ | Gobat et al. 2012 |
| 22 | 39138 | 214.87123 | +52.84507 | $22.13 \pm 0.04^\spadesuit$ | 3.442 | $10.95^{+0.03}_{-0.03}$ | Jin et al. 2024 |
| 23 | 1047519 | 150.612879 | +2.41102 | 23.16 | 3.442 | $10.71^{+0.04}_{-0.04}$ | Kakimoto et al. 2024 |
| 24 | J221724.8+001803.7 | 334.353333 | +0.3010278 | 22.3 | $3.3868^{+0.0010}_{-0.0010}$ | $11.45^{+0.35}_{-0.17}$ | Kubo et al. 2015 |
| 25 | J221737.3+001823.2 | 334.405417 | +0.3064444 | 22.5 | $3.0851^{+0.0001}_{-0.0001}$ | $10.90^{+0.23}_{-0.20}$ | Kubo et al. 2015 |
| 26 | J221732.0+001655.5 | 334.383333 | +0.2820833 | 22.2 | $3.0909^{+0.0004}_{-0.0004}$ | $11.06^{+0.49}_{-0.24}$ | Kubo et al. 2015 |
| 27 | J221737.3+001630.7 | 334.4054167 | +0.2751944 | 22.0 | $3.0888^{+0.0004}_{-0.0004}$ | $10.91^{+0.00}_{-0.07}$ | Kubo et al. 2015 |
| 28 | J221725.4+001716.9 | 334.3558333 | +0.2880278 | 21.7 | $3.0482^{+0.0003}_{-0.0003}$ | $11.05^{+0.06}_{-0.09}$ | Kubo et al. 2015 |
| 29 | ADF22-QG1 | 334.4052083 | +0.3044444 | 22.58 | $3.0922^{+0.008}_{-0.004}$ | $10.99^{+0.07}_{-0.08}$ | Kubo et al. 2022 |
| 30 | m1423 | 334.4052083 | +0.3044444 | $20.58\dagger$ | $3.2092^{+0.0002}_{-0.0002}$ | $10.58^{+0.08}_{-0.08}$ | Man et al. 2021 |
| 31 | 3D-EGS-18996 | 214.89563 | +52.856556 | $21.60 \pm 0.02$ | $3.250^{+0.002}_{-0.002}$ | $10.88^{+0.00}_{-0.04}$ | Nanayakkara et al. 2024, Schreiber et al. 2018, Esdaile et al. 2021 |
| 32 | 3D-EGS-31322 | 214.86606 | +52.884312 | $22.20 \pm 0.04$ | $3.434^{+0.001}_{-0.001}$ | $10.88^{+0.00}_{-0.04}$ | Nanayakkara et al. 2024, Schreiber et al. 2018, Esdaile et al. 2021, Jin et al. 2024 |
| 33 | 3D-EGS-34322 | 214.81316 | +52.858986 | $23.52 \pm 0.21$ | $3.227^{+0.004}_{-0.004}$ | $10.16^{+0.03}_{-0.01}$ | Nanayakkara et al. 2024, Schreiber et al. 2018 |
| 34 | ZF-UDS-3651 | 34.289452 | −5.2698030 | $22.95 \pm 0.03$ | $3.813^{+0.001}_{-0.000}$ | $10.65^{+0.01}_{-0.01}$ | Nanayakkara et al. 2024, Schreiber et al. 2018 |
| 35 | ZF-UDS-4347 | 34.290428 | −5.2620687 | $23.17 \pm 0.03$ | $3.703^{+0.002}_{-0.001}$ | $10.45^{+0.00}_{-0.03}$ | Nanayakkara et al. 2024, Schreiber et al. 2018 |
| 36 | ZF-UDS-6496 | 34.340358 | −5.2412550 | $22.62 \pm 0.02$ | $3.976^{+0.002}_{-0.002}$ | $10.86^{+0.00}_{-0.02}$ | Nanayakkara et al. 2024, Schreiber et al. 2018 |
| 37 | ZF-UDS-7329 | 34.255872 | −5.2338210 | $22.36 \pm 0.01$ | $3.207^{+0.002}_{-0.001}$ | $11.10^{+0.01}_{-0.04}$ | Nanayakkara et al. 2024, Schreiber et al. 2018, Glazebrook et al. 2023 |
| 38 | ZF-UDS-7542 | 34.258888 | −5.2322803 | $23.02 \pm 0.02$ | $3.208^{+0.000}_{-0.002}$ | $10.69^{+0.01}_{-0.03}$ | Nanayakkara et al. 2024, Schreiber et al. 2018 |
| 39 | ZF-UDS-8197 | 34.293755 | −5.2269478 | $23.20 \pm 0.03$ | $3.550^{+0.000}_{-0.001}$ | $10.40^{+0.01}_{-0.01}$ | Nanayakkara et al. 2024, Schreiber et al. 2018 |
| 40 | 3D-UDS-35168 | 34.485131 | −5.1578340 | $23.89 \pm 0.09$ | $3.529^{+0.007}_{-0.004}$ | $10.18^{+0.01}_{-0.02}$ | Nanayakkara et al. 2024, Schreiber et al. 2018 |
| 41 | 3D-UDS-39102 | 34.526210 | −5.1438100 | $23.28 \pm 0.15$ | $3.587^{+0.000}_{-0.001}$ | $10.77^{+0.08}_{-0.02}$ | Nanayakkara et al. 2024, Schreiber et al. 2018 |
| 42 | 3D-UDS-41232 | 34.526589 | −5.1360390 | $21.74 \pm 0.01$ | $3.121^{+0.001}_{-0.002}$ | $11.17^{+0.00}_{-0.01}$ | Nanayakkara et al. 2024, Schreiber et al. 2018 |
| 43 | ZF-COS-20115 | 150.06149 | +2.3787093 | $22.43 \pm 0.02$ | $3.715^{+0.002}_{-0.002}$ | $11.06^{+0.16}_{-0.09}$ | Schreiber et al. 2018, Glazebrook et al. 2017, Esdaile et al. 2021 |
| 44 | ZF-COS-20133 | 150.12173 | +2.3745940 | $23.84 \pm 0.06$ | $3.481^{+0.0002}_{-0.0002}$ | $10.52^{+0.01}_{-0.06}$ | Schreiber et al. 2018 |
| 45 | 3D-EGS-26047 | 214.90512 | +52.891621 | $22.55 \pm 0.05$ | $3.234^{+0.0020}_{-0.0016}$ | $11.00^{+0.21}_{-0.14}$ | Schreiber et al. 2018 |
| 46 | 3D-EGS-40032 | 214.76062 | +52.845383 | $21.59 \pm 0.03$ | $3.219^{+0.0011}_{-0.0013}$ | $11.31^{+0.16}_{-0.14}$ | Schreiber et al. 2018, Esdaile et al. 2021 |
| 47 | ZF-COS-17779 | 150.04651 | +2.3673911 | $23.82 \pm 0.06$ | $3.415^{+0.1320}_{-0.0003}$ | $10.56^{+0.12}_{-0.16}$ | Schreiber et al. 2018 |
| 48 | ZF-COS-18842 | 150.08728 | +2.3960431 | $23.03 \pm 0.04$ | $3.782^{+0.0023}_{-0.0031}$ | $10.65^{+0.06}_{-0.04}$ | Schreiber et al. 2018 |
| 49 | ZF-COS-19589 | 150.06671 | +2.3823645 | $23.54 \pm 0.06$ | $3.715^{+0.0094}_{-0.1589}$ | $10.79^{+0.10}_{-0.08}$ | Schreiber et al. 2018 |

**Table 3** *(continued)*

| | Galaxy ID | $\alpha$ | $\delta$ | $K_s$ (AB mag) | $z_{spec}$ | $\log(M_*/M_\odot)$ | References |
|---|---|---|---|---|---|---|---|
| | (1) | (2) | (3) | (4) | (5) | (6) | (7) |
| 50 | UNCOVER 18407 | 3.551866 | −30.392230 | $25.36 \pm 0.01^\S$ | $3.970^{+0.0006}_{-0.0006}$ | $10.38^{+0.08}_{-0.09}$ | Setton et al. 2024 |
| 51 | SXDS2_19838 | 34.382500 | −5.3396111 | 22.72 | $3.985^{+0.003}_{-0.003}$ | $11.05^{+0.09}_{-0.61}$ | Tanaka et al. 2023[†] |
| 52 | SXDS2_15659 | 34.380833 | −5.3833333 | 23.46 | $3.993^{+0.005}_{-0.030}$ | $10.81^{+0.05}_{-0.05}$ | Tanaka et al. 2023[†] |
| 53 | SXDS2_16609 | 34.406667 | −5.3734444 | 23.99 | $3.687^{+0.271}_{-0.081}$ | $10.46^{+0.08}_{-0.07}$ | Tanaka et al. 2023[†] |
| 54 | SXDS2_19229 | 34.435833 | −5.3457222 | 23.74 | $4.009^{+0.009}_{-0.158}$ | $10.89^{+0.04}_{-0.07}$ | Tanaka et al. 2023[†] |
| 55 | SXDS2_19997 | 34.446250 | −5.3386667 | 23.67 | $3.993^{+0.005}_{-0.423}$ | $10.74^{+0.04}_{-0.09}$ | Tanaka et al. 2023[†] |
| 56 | SXDS-10017 | 34.75625 | −5.30804 | 22.5 | $3.767^{+0.103}_{-0.001}$ | $10.89^{+0.05}_{-0.06}$ | Valentino et al. 2020 |
| 57 | SXDS-27434 | 34.29871 | −4.98987 | 21.9 | $4.013^{+0.0005}_{-0.0005}$ | $11.06^{+0.04}_{-0.04}$ | Valentino et al. 2020, Tanaka et al. 2023 |
| 58 | COS-466654 | 149.41958 | +2.00755 | 22.3 | $3.775^{+0.002}_{-0.003}$ | $10.82^{+0.03}_{-0.03}$ | Valentino et al. 2020, Tanaka et al. 2019 |

NOTE—Coordinates, $K_s$ magnitudes, and stellar masses of the 58 spectroscopically-confirmed quiescent galaxies with $\log(M_*/M_\odot) > 10$ at $z > 3$ in the literature. (4) Where not provided in the listed reference(s), we list the $K_s$ magnitudes estimated by cross-matching using publicly-available catalogs or by integrating the spectrum over the $K_s$ band. (7) We list references that present spectroscopy of the object(s) and the respective photometric selection paper if the coordinates and/or $K_s$ magnitudes are not provided in the former.

[†] Spectrophotometric redshifts, i.e., derived by jointly fitting the photometry and spectroscopy.

[◇] Derived by cross-matching galaxies with the COSMOS2020 catalog Weaver et al. (2022)

[♠] Derived by cross-matching galaxies with 3D-HST catalogs Skelton et al. (2014)

[♣] Estimated from the galaxy's spectrum in JADES DR3[5] with uncertainties derived from the NIRSpec error spectrum.

[*] Estimated from the $K_s$ photometry provided in Figure 2 of Gobat et al. (2012).

[‡] Based on a $K$-correction to the F160W magnitude provided in Table 8 of Man et al. (2021) i.e., $K_s = F160W − 2.5 \log(1 + z)$ (Blanton & Roweis 2007).

[§] Estimated from the galaxy's spectrum with uncertainties derived from the full posterior of the fitted model (*priv. communication with authors*).

## ACKNOWLEDGEMENTS

This work benefited from the generous support of the George P. and Cynthia Woods Mitchell Institute for Fundamental Physics and Astronomy at Texas A&M University. The Gemini/Flamingos-2 images presented in this work were reduced on the OzSTAR supercomputing facility at Swinburne University of Technology. Portions of this research were conducted with the advanced computing resources provided by Texas A&M High Performance Research Computing. This material is based upon work supported by the National Science Foundation under grants AST-2009632 and AST-2009442.

Based on observations obtained at the international Gemini Observatory, a program of NSF's OIR Lab, which is managed by the Association of Universities for Research in Astronomy (AURA) under a cooperative agreement with the National Science Foundation on behalf of the Gemini Observatory partnership: the National Science Foundation (United States), National Research Council (Canada), Agencia Nacional de Investigación y Desarrollo (Chile), Ministerio de Ciencia, Tecnología e Innovación (Argentina), Ministério da Ciência, Technologia, Inovações e Comunicações (Brazil), and Korea Astronomy and Space Science Institute (Republic of Korea).

Based on data products from observations made with ESO Telescopes at the La Silla Paranal Observatory under ESO programme ID 179.A-2005 and on data products produced by CALET and the Cambridge Astronomy Survey Unit on behalf of the UltraVISTA consortium.

This research uses observations made with the NASA/ESA Hubble Space Telescope (associated with program GO-14114), obtained at the Space Telescope Science Institute, which is operated by the Association of Universities for Research in Astronomy, Inc., under NASA contract NAS 5-26555.

This research has made use of the Keck Observatory Archive (KOA), which is operated by the W. M. Keck Observatory and the NASA Exoplanet Science Institute (NExScI), under contract with the National Aeronautics and Space Administration. This paper is based primarily on observations taken at the W. M. Keck Observatory and we acknowledge the important cultural role that the summit of Mauna Kea has within the indigenous Hawaiian community.

K. Tran acknowledges support by the Australian Research Council Centre of Excellence for All Sky Astrophysics in 3 Dimensions (ASTRO 3D), through project number CE170100013. The authors would like to thank our Keck Observatory Support Astronomer Chien-Hsui Lee for assistance during our observing run. Special thanks to our Operating Assistant, Tony Ridenour for driving the telescope during our observations. We would also like to thank Nick Martis for providing us with UltraVISTA catalogs supplemented with far-IR data and the accompanying stellar population parameters from MAGPHYS.

The synthetic $(ugi)_s$ filters are available on Github[6]. The Kurucz stellar models are available via the STScI archive [7]. The raw images for the FENIKS Survey (under program ID #) are available from the Gemini archive[8]. We plan to release the catalogs with a future paper outlining the data reduction and catalog procedure for the survey (Antwi-Danso et al. in prep). Other data products can be provided upon request to the corresponding author.

*Software:* `Astropy` (Astropy Collaboration et al. 2018), `Bagpipes` (Carnall et al. 2018), `eazy-py` (Brammer 2021), `Jupyter` (Kluyver et al. 2016), `matplotlib` (Hunter 2007), `numpy` (Harris et al. 2020), `pandas` (The Pandas Development Team 2020), `SEP` (Barbary 2016) `seaborn` (Waskom 2021), `scipy` (Virtanen et al. 2020), `lmfit` (Seabold & Perktold 2010)

*Facility:* Gemini:South, Keck:I (MOSFIRE)

## APPENDIX

### A. OPTIMAL EXTRACTION, FLUX CALIBRATION AND TELLURIC CORRECTION

Here, we describe the procedure used to collapse the reduced 2D MOSFIRE spectra (Section 2.2) into 1D spectra by summing along the wavelength axis. We followed the extraction technique outlined in Horne (1986), which enhances the signal-to-noise ratio (S/N) of detected sources by weighting the spectra by the inverse variance and the expected spatial profile of the source. This ensures that pixels far from the peak of the detected source are assigned lower weight, as they receive little light from the source. To account for variations in seeing as a function of wavelength, we applied the 2D spatial profile of a slit star in each mask to the corresponding source spectra prior to collapsing the spectra (see also Song et al. 2016). The 2D errors from the MOSFIRE DRP are weighted similarly and summed in quadrature. Figure 8 shows a slit star in one of our four masks extracted using uniform (boxcar) extraction versus the optimized extraction method. We see that this method increases the S/N of the extracted 1D spectrum quite appreciably, by a factor of $\sim 2$.

This optimized extraction technique depends on accurate knowledge of the spatial distribution of the source light as a function of wavelength. Sky lines and hot pixels can produce secondary peaks in the spatial profile of detected objects. For the faintest sources ($K_s < 23$), these secondary bumps may have a slightly higher amplitude than that of the true spatial profile. This was the case for our observations of COS55-126981 ($z_{spec} = 4.673$) in mask `FENIKS_COSMOS55_22A_4`. We isolated the true peak by masking pixels with values greater than 10% of the median error corresponding to those observations. Because `FENIKS_COSMOS55_22A_4` has the same exposure time and slit set up as `FENIKS_COSMOS55_22A_1_1`, we used the latter to obtain an estimate of the true spatial profile and excluded pixels containing the object from the sigma-clipping procedure.

Figure 9 shows the spatial profile of a secure (median S/N > 3/pixel) detection (COS55-128636) and a tentative detection (COS55-126981). In both cases, the profile exhibits a peak characteristic of a detected source, and the troughs corresponding to the negative traces created by the dither pattern (Section 2). We contrast the latter with a 1D extraction from a blank region of the 2D spectrum to emphasize that although COS55-126981 is faint, it is detected.

Following Nanayakkara et al. (2016), we observed a standard star at the start and end of each night to be used for flux calibration and to correct for telluric absorption. Unfortunately, our standard star observations were truncated during our runs, rendering them unusable. Each of our masks contained at least one "slit star," which is observed contemporaneously with our targets, and hence can be used to monitor variations in seeing and transmission. We first determined the spectral type of each slit star by estimating its rest-frame $B - J$ and $J - K$ colors (Epchtein et al. 1994), which have been shown to delineate stars of various spectral classes into a clear sequence. To determine the luminosity class of the slit star, we fit a Blackbody to its UltraVISTA and FENIKS photometry. Figure 10 summarizes this process. We then use the corresponding Kurucz stellar spectrum, which is provided in physical (cgs) units, for our flux calibration and telluric correction. The best-fit stellar spectrum is scaled to the photometry of the slit star.

We determined the flux calibration for each mask by determining the continuum of the model spectrum and that of the slit star spectrum using a spline fit. The calibration is then computed as the ratio of the two. The telluric correction is then computed as the ratio of the continuum fit to the model spectrum (intrinsic) and the observed spectrum. We illustrate this for a G0I slit star in Figure 11. Because slit stars tend to be fainter than standards, their S/N is lower, resulting in noisier telluric corrections. This can degrade the S/N of our targets after the telluric correction is applied. To mitigate this, we create a stacked telluric correction from all masks, which has much higher S/N.

---

[6] https://github.com/jacqdanso/synthetic-ugi-filters

[7] https://www.stsci.edu/hst/instrumentation/reference-data-for-calibration-and-tools/astronomical-catalogs/kurucz-1993-models

[8] https://archive.gemini.edu/searchform/

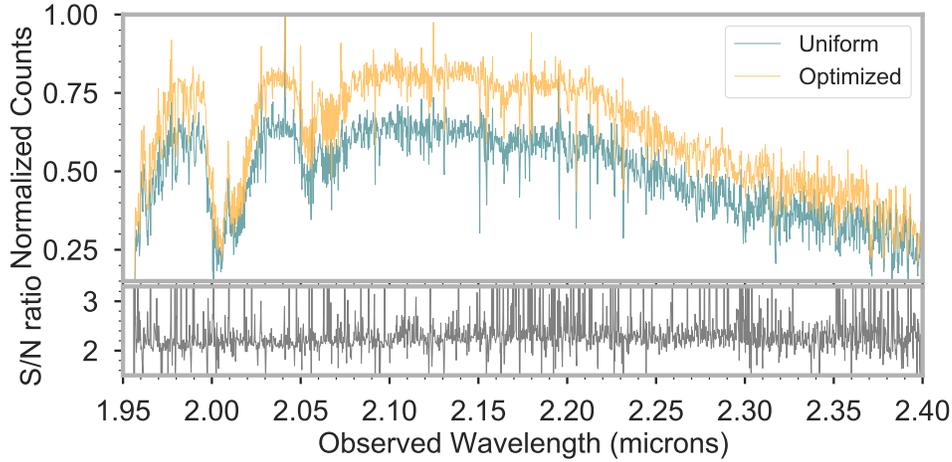

**Figure 8.** Uniform (boxcar) extraction versus optized extraction for the slit star in mask `FENIKS_COSMOS55_22A_1_1`. Because this star is relatively faint ($K_s$ = 19.4 AB), its S/N is increased quite substantially via optimized extraction (by a factor of ~ 2).

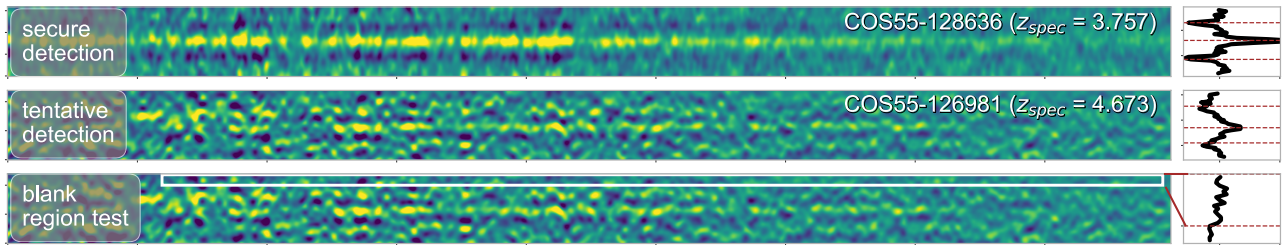

**Figure 9.** Blank region test for COS55-126981 ($z_{spec}$ = 4.673). **Left:** Coadded 2D spectra from Figure 3. **Right:** Flux summed in the spatial direction (along the rows of the 2D spectrum) and normalized by the peak flux of COS55-128636. The dotted lines mark the regions used for optimized extraction of the 1D spectra. The profile of COS55-126981 has the shape characteristic of a continuum detection, although it has a lower peak than COS55-128636 due to lower S/N. The spatial profile of a blank region on the bottom panel confirms this, as it is flat compared to the previous two, indicating the absence of an object. Similar to COS55-126981, the S/N of the second-faintest object (COS55-129098) was sufficient to determine its spatial profile for 1D extraction.

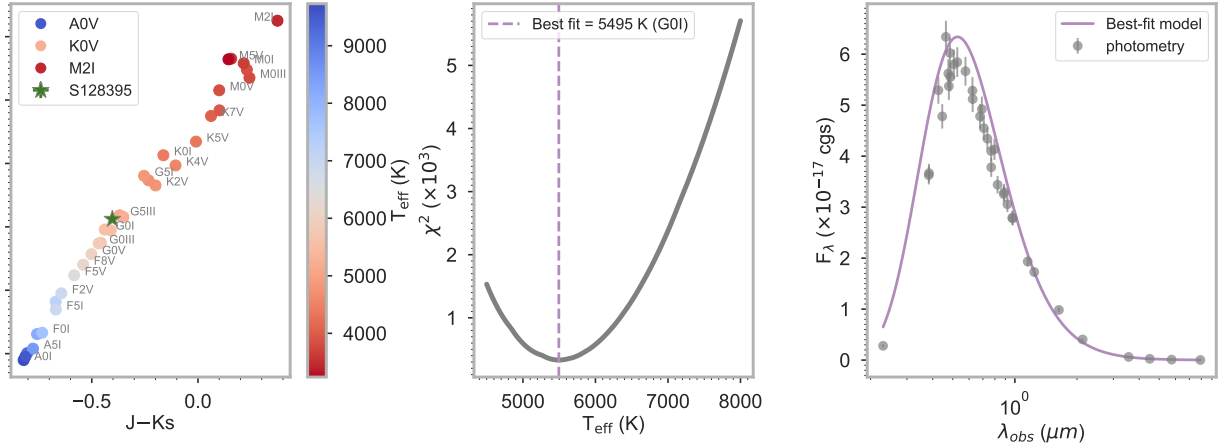

**Figure 10.** Procedure used for determining the spectral types of our slit stars. Using the star in Figure 8 as an example, we obtained an intial estimate of the spectral class by comparing the $B - J$ and $J - K_s$ colors of our slit stars to those of Kurucz stellar models (**Left**). We determined the luminosity class by fitting a Blackbody curve to the star's photometry (**Middle** and **Right**). The offset between the best-fit model and the photometry is due to Balmer and metal absorption features in the star's spectrum.

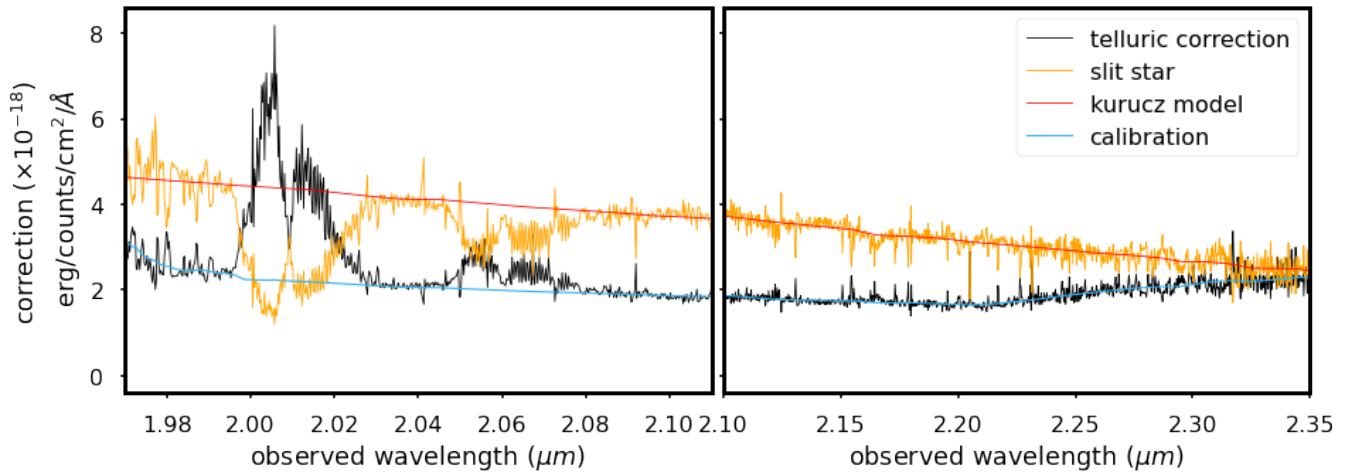

**Figure 11.** Transmission correction for the example slit star in Figure 8 (a G0I star) in mask `FENIKS_COSMOS55_22A_1_1`. The correction takes into account the absolute flux calibration and telluric absorption.